\documentclass[12pt]{article}
 
\usepackage[margin=1in]{geometry} 
\usepackage{parskip}
\usepackage{setspace}
\usepackage{amsmath,amsthm,amssymb}
\usepackage{placeins}
\usepackage{float}
\usepackage{enumerate}
\usepackage{graphicx}
\usepackage{xcolor}
\usepackage{array, booktabs, longtable, multirow}
\usepackage[round,authoryear]{natbib}
\usepackage[colorlinks=true,citecolor=blue]{hyperref}
\newtheorem{theorem}{Theorem}
\newtheorem{assumption}{Assumption}

\newtheorem{proposition}{Proposition}
\usepackage[font=small,labelfont=bf,labelsep=period]{caption}
\usepackage{authblk}
\usepackage{tikz}
\usetikzlibrary{arrows.meta, positioning}

\usepackage{etoolbox}

\makeatletter
\newcommand{\printappendixtoc}{%
  \section*{Appendix Contents}
  \begingroup
    \setcounter{tocdepth}{2}
    \@starttoc{apc}
  \endgroup
}
\makeatother

\newcommand{\appsection}[1]{%
  \section{#1}%
  \addcontentsline{apc}{section}{\protect\numberline{\thesection}#1}%
}
\newcommand{\appsubsection}[1]{%
  \subsection{#1}%
  \addcontentsline{apc}{subsection}{\protect\numberline{\thesubsection}#1}%
}

\newcommand{\indep}{\perp \!\!\! \perp}

\title{Proximal Reconciliation of Conflicting Trials with Unmeasured Unblinding}
\author[1]{Daniel A Xu}
\author[1]{Antonio Olivas-Martinez}
\author[2]{Brandie D Taylor}
\author[3]{Beth L Pineles}
\author[4]{Sean C Blackwell}
\author[1,3]{Sunni L Mumford}
\author[3]{Kristin D Gerson}
\author[1]{Enrique F Schisterman}
\author[1]{Eric J Tchetgen Tchetgen}
\author[1]{Ellen C Caniglia}

\affil[1]{Department of Biostatistics, Epidemiology, and Informatics, Perelman School of Medicine, University of Pennsylvania}
\affil[2]{Obstetrics and Gynecology, The University of Texas Medical Branch}
\affil[3]{Department of Obstetrics \& Gynecology, Perelman School of Medicine, University of Pennsylvania}
\affil[4]{Obstetrics and Gynecology, Children’s Memorial Hermann Hospital}

\begin{document}

\maketitle

\begin{abstract}
    The conflicting findings from the Meis and PROLONG trials have raised questions about the effect of 17-alpha-hydroxyprogesterone caproate (17OHP-C) on preterm birth (delivery before 37 weeks gestation). We investigated whether differences in treatment effects mediated through unblinding contributed to these conflicting results. We propose a framework for assessing reconciliation by comparing ``direct" treatment effects across trials. Because unblinding was unmeasured, we used injection site reactions and medication prescriptions as proxies and applied nonparametric proximal causal mediation methods to estimate the natural direct and indirect effects of 17OHP-C on preterm birth. We also estimated controlled direct effects under a hypothetical intervention eliminating injection site reactions and conducted sensitivity and subgroup analyses. The estimated natural direct effect was $-19.2\%$ in Meis and $0.1\%$ in PROLONG, a difference of $-19.3\%$ (95\% confidence interval: $-28.4\%, -10.1\%$). The controlled direct effect was $-18.0\%$ in Meis and $-0.3\%$ in PROLONG, a difference of $-17.7\%$ (95\% confidence interval: $-31.6\%, -3.7\%$). Further accounting for differences in baseline covariate distributions between trials using novel transportability-based mediation estimators provided no definitive evidence of greater reconciliation. These findings suggest that differences in the effects of hidden unblinding, as captured by the proposed proxies, appear insufficient to reconcile the conflicting findings.
\end{abstract}

\section{Introduction}
\label{s:intro}
In 2003, a multicenter, randomized, double-blind, placebo-controlled trial conducted by \citet{meis2003prevention} reported that 17-alpha-hydroxyprogesterone caproate (17OHP-C) substantially reduced the risk of recurrent preterm birth among pregnant individuals with a prior spontaneous preterm birth. The trial estimated a risk difference (RD) of $-18.6\%$ (95\% confidence interval [CI]: $-28.2$\%, $-9.2$\%) for delivery before 37 weeks’ gestation, including spontaneous abortions and stillbirths \citep{meis2003prevention}. Based on these findings, the trial was stopped early for benefit, and 17OHP-C subsequently received accelerated approval from the U.S. Food and Drug Administration. However, the subsequent Progestin’s Role in Optimizing Neonatal Gestation (PROLONG) trial, which was designed to confirm the earlier findings using the same treatment regimen and nearly identical eligibility criteria, did not demonstrate evidence that 17OHP-C reduced the risk of recurrent preterm birth (RD: 1.2\%, 95\% CI: $-3.0$\%, $5.3$\%) \citep{blackwell202017}.

The conflicting findings from the Meis and PROLONG trials have generated substantial clinical and regulatory uncertainty, especially as preterm birth remains a leading cause of neonatal morbidity and mortality. Several explanations for the conflicting findings have been proposed, including differences in the underlying study populations, differential nonadherence, and chance. However, analyses investigating some of these hypotheses -- in particular, the role of differences in the distributions of measured and unmeasured effect modifiers across the two trial populations -- have failed to reconcile the findings between the two trials, suggesting that other explanations may be responsible \citep{virkud2025clarifying, xu2026regression}.

Another potential explanation that warrants further investigation is that the two trials differed with respect to post-randomization protocol deviations. In double-blind randomized trials, one such deviation is unplanned unblinding, in which participants, investigators, or other study personnel unintentionally become aware of the assigned treatment \citep{evans2010fundamentals}. If treatment assignment influences the risk of unblinding, and patients’ or clinicians’ perceptions of treatment assignment subsequently influence clinical management, treatment adherence, or other downstream behaviors related to the risk of preterm birth, then unblinding may mediate the effect of treatment assignment on the outcome. In the 17OHP-C trials, indirect evidence supporting this hypothesis comes from differences in the prevalence of reported injection site reactions between the two studies and by randomization arm, as these reactions represent one mechanism by which treatment assignment may become unblinded. For example, injection site pain was more commonly reported in Meis than in PROLONG (151/443 [34.1\%] vs. 70/1,676 [4.2\%], respectively, Table \ref{tab:table1}). Additionally, injection site reactions other than pain, including swelling, nodules, and pruritus, were more common among participants randomized to 17OHP-C than among those randomized to placebo, especially in Meis. For instance, 53/171 (31.0\%) participants assigned to 17OHP-C in Meis reported injection site swelling, compared with 12/81 (14.8\%) assigned to placebo (Appendix Table \ref{tab:injection-site-reactions}).

Under this hypothesis, unblinding can be viewed as a mediator on the causal pathway from treatment assignment to preterm birth. Since the seminal work of \citet{baron1986moderator} and the introduction of natural direct and indirect effects \citep{robins1992identifiability, pearl2001direct}, causal mediation analyses, which provide a framework for investigating the mechanisms through which an exposure affects an outcome, have been studied extensively. In our setting, the natural indirect effect (NIE) captures the portion of the treatment effect operating through unblinding and subsequent changes in patient or clinician behavior that may affect preterm birth risk, whereas the natural direct effect (NDE) captures the remaining portion of the treatment effect operating through pathways other than unblinding (Figure \ref{fig:dags}). Differences between Meis and PROLONG in either the effect of treatment assignment on unblinding or the effect of unblinding on the outcome could have contributed to the trials’ conflicting findings by altering the decomposition of the total effect into direct and indirect effects. For example, a substantially attenuated NDE in Meis, relative to its total effect, might suggest that a meaningful portion of the observed treatment effect operated through the unblinding pathway rather than through a direct effect on the risk of preterm birth. Conversely, in PROLONG, a protective NDE could have been offset by a harmful NIE, resulting in little or no overall treatment effect \citep{mackinnon2000equivalence}. In such settings, the total effect alone may obscure important underlying mechanisms. Comparing the natural direct and indirect effects across the two trials may therefore provide insight into whether differential unblinding contributed to their conflicting findings.

Standard identification of the NDE and NIE relies on strong cross-world independence assumptions, which generally fail in the presence of treatment-induced mediator-outcome confounding, even when the relevant confounders are measured \citep{vanderweele2014effect}. Therefore, alternative estimands could also be considered to characterize the role of unblinding. One alternative is the controlled direct effect (CDE), which contrasts treatment assignments while fixing the mediator to a specified value \citep{robins1992identifiability}. The CDE can be identified without cross-world independence assumptions and addresses questions such as: what would be the effect of assignment to 17OHP-C versus placebo under a hypothetical intervention preventing unblinding in all participants? Alternatively, randomized interventional direct effects (RIDEs) and indirect effects (RIIEs) replace each individual’s natural mediator value with a random draw from the mediator distribution under a specified treatment assignment, conditional on baseline covariates. These effects characterize treatment effects under interventions on the distribution of unblinding and likewise avoid cross-world independence assumptions \citep{vanderweele2014effect, vansteelandt2017interventional, diaz2021nonparametric, zheng2017longitudinal, rudolph2017robust, nguyen2021clarifying}. Comparing these alternative estimands across trials could provide complementary perspectives on whether differential unblinding can reconcile conflicting trial findings.

A key challenge is that unblinding was not directly measured in either trial, thereby precluding the use of standard mediation analyses. Recent work on proximal causal inference provides a framework for identifying natural direct and indirect effects when the mediator itself is unmeasured but informative proxy variables are available \citep{ghassami2025causal}. This framework leverages two types of proxy variables: one that captures the pathway from treatment assignment to unblinding, and another that captures the pathway from unblinding to preterm birth. In our application, we propose injection site reactions as the former proxy and prescriptions of tocolytics, corticosteroids, and/or antibiotics related to preterm birth as the latter. Injection site reactions may provide a mechanism through which treatment assignment leads to unblinding, whereas the aforementioned medication prescriptions may reflect resulting changes in clinical management due to perceived treatment assignment that affect preterm birth risk. More formally, these proxies must satisfy the conditional independence assumptions in Section \ref{s:statisticalanalysis}.

In this paper, we apply existing nonparametric proximal hidden mediation methods to the Meis and PROLONG trials to investigate whether differential unblinding, i.e., differences between trials in the effect of treatment assignment on unblinding and/or the effect of unblinding on the outcome, may help reconcile the conflicting findings. Within each trial, we estimate the total effect of assignment to 17OHP-C versus placebo, the NDE not operating through unblinding, and the NIE operating through unblinding. We then compare these effects across trials to assess reconciliation. As a complementary analysis, we also estimate and compare the CDEs under a hypothetical intervention eliminating injection site reactions. We conduct sensitivity and subgroup analyses to evaluate the robustness of our findings. In Appendix \ref{app:A-additional}, we develop novel transportability-based estimators that extend these analyses to account for differences in covariate distributions between the trials, and in Appendix \ref{app:appendixB}, we propose novel identification results for randomized interventional mediation effects when the mediator is unmeasured, which may be of independent interest. More broadly, this work provides the first formal evaluation of the hypothesis that differential unblinding contributed to the conflicting findings of the Meis and PROLONG trials while illustrating the practical application of recently developed methods for causal mediation analysis with hidden mediators.

\section{Methods}
\label{s:methods}

\subsection{Data Source and Study Population}
\label{s:population}

We analyzed de-identified participant-level data from the Meis and PROLONG trials, obtained from AMAG Pharmaceuticals, Inc. Meis was a multicenter, randomized, double-blind, placebo-controlled trial conducted in the United States among 463 pregnant individuals \citep{meis2003prevention}. PROLONG was a randomized, double-blind, placebo-controlled confirmatory trial that enrolled 1,708 pregnant individuals across 93 centers in 9 countries \citep{blackwell202017}. In both trials, eligible participants had a history of spontaneous preterm birth and were randomized to receive weekly intramuscular injections of 17OHP-C or placebo beginning between 16 weeks and 20 weeks and 6 days of gestation and continuing until delivery or 36 weeks' gestation. Injections were administered by research staff independently of treating clinicians.

The primary outcome for our analysis was recurrent preterm birth, defined as delivery before 37 weeks’ gestation. Although delivery before 37 weeks was the primary efficacy outcome in Meis and a secondary efficacy outcome in PROLONG, we used this common outcome definition to allow for direct comparison across trials. Treatment was defined as randomized assignment to 17OHP-C versus placebo.

Participants with missing outcome data (4 in Meis and 24 in PROLONG) or missing baseline covariates (16 in Meis and 8 in PROLONG) were excluded from the primary analysis, resulting in analytic samples of 443 participants from Meis and 1,676 participants from PROLONG.

\subsection{Statistical Analysis}
\label{s:statisticalanalysis}
Let $S\in\{0,1\}$ denote the trial in which an individual participated (1: Meis, 0: PROLONG). Let $A \in \{0,1\}$ denote randomized treatment assignment (1: 17OHP-C, 0: placebo). Let $Y$ denote the binary indicator of delivery before 37 weeks gestation (1: yes, 0: no). Let $X$ denote the measured baseline covariates summarized in Table \ref{tab:table1}, including but not limited to maternal age, maternal race, and gestational age at qualifying prior preterm birth. Let $M$ denote a binary indicator of unmeasured unblinding (1: yes, 0: no). Finally, let $Z$ and $W$ denote observed proxy variables for $M$,  described in greater detail below.

Let $Y(a,m)$ denote the potential outcome that would be observed under treatment assignment $a$ and unblinding status $m$, and let $M(a)$ denote the unblinding status that would occur under treatment assignment $a$. Denote the counterfactual mean outcome in trial $s$ as $\psi_s(a,a') = \mathbb{E}\{Y(a,M(a'))\mid S=s\}$, which represents the mean outcome if individuals were assigned treatment $a$, while their unblinding status were set to the value it would have attained under a potentially different treatment assignment $a'$. The total effect in each trial $s$ is defined as
$$\text{ATE}_s = \psi_s(1,1)-\psi_s(0,0).$$
The ATE in each trial can be expressed as the sum of the NDE and NIE, where the NDE in trial $s$ is
$$\text{NDE}_s = \psi_s(1,0)-\psi_s(0,0)$$
and the NIE in trial $s$ is
$$\text{NIE}_s = \psi_s(1,1)-\psi_s(1,0).$$
The NDE contrasts assignment to 17OHP-C versus placebo while fixing unblinding status to the level it would naturally attain under placebo, whereas the NIE contrasts the level of unblinding that would naturally occur under 17OHP-C versus placebo while holding treatment assignment fixed at 17OHP-C. The NDE and NIE can alternatively be defined using $\psi_s(0,1)$ rather than $\psi_s(1,0)$; corresponding analyses based on $\psi_s(0,1)$ are presented in Appendix \ref{app:A-additional}, with minimal impact on the overall conclusions.

In the primary analysis, the proxy $Z$ was a two-dimensional vector indicating the occurrence of injection site pain and at least one injection site reaction other than pain (e.g., swelling, nodules, or pruritus) at any visit. The proxy $W$ was defined as a three-dimensional vector indicating prescriptions for tocolytics, corticosteroids, and antibiotics for indications related to preterm birth (e.g., premature rupture of membranes, fetal lung maturation, infant respiratory distress syndrome). Appendix \ref{app:A-measurement} describes proxy assessment by trial and lists all reported injection site reactions.

\begin{figure}[htbp]
\centering

\resizebox{\textwidth}{!}{%
\begin{tikzpicture}[
    >=Latex,
    node distance=1.1cm and 1.35cm,
    nodebox/.style={
        draw,
        rounded corners=5pt,
        minimum width=2.45cm,
        minimum height=1.30cm,
        align=center,
        font=\large,
        inner sep=4pt
    },
    causal/.style={
        ->,
        line width=1.1pt,
        draw=red
    },
    covariate/.style={
        ->,
        line width=1.1pt,
        draw=black
    },
    direct/.style={
        ->,
        line width=1.1pt,
        draw=blue!65!black
    }
]

\begin{scope}[local bounding box=leftdag]

\node[nodebox] (A1) {
    \textbf{A}\\
    17OHP-C
};

\node[nodebox, right=of A1] (M1) {
    \textbf{M}\\
    Unblinding
};

\node[nodebox, right=of M1] (Y1) {
    \textbf{Y}\\
    Preterm Birth
};

\node[nodebox, below=1.15cm of M1] (X1) {
    \textbf{X}\\
    Covariates
};

\draw[causal] (A1) -- (M1);
\draw[causal] (M1) -- (Y1);

\draw[direct]
    (A1.north)
    to[out=20,in=160,looseness=1.05]
    (Y1.north);

\draw[covariate]
    (X1.north west)
    to[out=160,in=290]
    (A1.south east);
\draw[covariate] (X1.north) -- (M1.south);

\draw[covariate]
    (X1.north east)
    to[out=20,in=250]
    (Y1.south west);

\end{scope}

\begin{scope}[
    xshift=12.5cm,
    local bounding box=rightdag
]

\node[nodebox] (A2) {
    \textbf{A}\\
    17OHP-C
};

\node[nodebox, right=of A2] (Z2) {
    \textbf{Z}\\
    Proxy
};

\node[nodebox, right=of Z2] (M2) {
    \textbf{M}\\
    Unblinding
};

\node[nodebox, right=of M2] (W2) {
    \textbf{W}\\
    Proxy
};

\node[nodebox, right=of W2] (Y2) {
    \textbf{Y}\\
    Preterm Birth
};

\node[nodebox, below=1.20cm of M2] (X2) {
    \textbf{X}\\
    Covariates
};

\draw[causal] (A2) -- (Z2);
\draw[causal] (Z2) -- (M2);
\draw[causal] (M2) -- (W2);
\draw[causal] (W2) -- (Y2);

\draw[causal]
    (A2.north east)
    to[out=25,in=155,looseness=1.05]
    (M2.north west);

\draw[causal]
    (M2.north east)
    to[out=25,in=155,looseness=1.05]
    (Y2.north west);

\draw[direct]
    (A2.north)
    to[out=25,in=155,looseness=1.25]
    (Y2.north);

\draw[covariate] (X2.west) -- (A2.south east);
\draw[covariate] (X2.north west) -- (Z2.south east);
\draw[covariate] (X2.north) -- (M2.south);
\draw[covariate] (X2.north east) -- (W2.south west);
\draw[covariate] (X2.east) -- (Y2.south west);

\end{scope}

\end{tikzpicture}%
}

\caption{
Directed acyclic graphs (DAGs) representing the hypothesized role of unblinding and proxy variables in each trial. Red arrows denote pathways measured by the natural indirect effect, and blue arrows denote pathways assessed by the natural direct effect. The arrow from $X$ to $A$ is included to illustrate the more general setting accommodated by the methods. When $A$ is randomized independent of baseline covariates (as in Meis and PROLONG), this arrow is absent.
}
\label{fig:dags}

\end{figure}
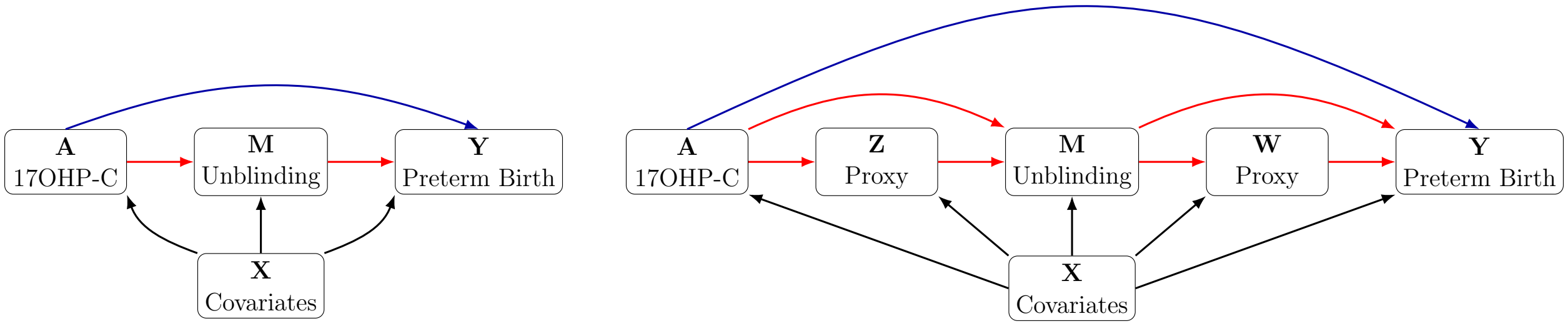

Identification of $\psi_s(a,a)$ (and therefore $\text{ATE}_s$) follows from standard consistency, positivity, and exchangeability assumptions (Appendix \ref{app:B-natural-identification}):
\begin{align*}
    \psi_s(a,a)=\mathbb{E}\bigl\{\mathbb{E}(Y\mid A=a,X,S=s)\mid S=s\bigr\}
\end{align*}
Identification of the NDE and NIE in each trial additionally leverages the proximal hidden mediation framework of \citet{ghassami2025causal}. This framework requires the standard mediation-specific consistency, positivity, and exchangeability assumptions (Appendix \ref{app:B-natural-identification}), including the aforementioned cross-world independence assumption \citep{tchetgen2012semiparametric, imai2010general}. Because the mediator is unobserved, identification in our setting also requires proxy conditions linking $Z$ and $W$ to the hidden mediator $M$. In particular, we assume that, within each trial, $Y \indep Z \mid (A,M,X, S=s)$ and $W \indep (A,Z) \mid (M,X,S=s)$. In our application, the first assumption would be satisfied if injection site reactions affect preterm birth only through unblinding and if there is no unmeasured confounding of the relationship between injection site reactions and preterm birth. Similarly, the second assumption would require that treatment assignment and injection site reactions affect the medication prescription proxy only through unblinding, and that there is no unmeasured confounding of the relationships between these variables and medication prescription. A directed acyclic graph (DAG) illustrating these assumptions is shown in Figure \ref{fig:dags}. Under these assumptions, together with completeness conditions and the existence of bridge functions $q_a(z,x)$ and $h_a(w,x)$ solving the integral equations in Appendix \ref{app:C-bridges},  identification of $\text{NDE}_s$ and $\text{NIE}_s$ follows from the identification of $\psi_s(a,a')$:
\begin{align*}
    \psi_s(a,a')
=\mathbb{E}\bigl[\mathbb{E}\{h_{s,a}(W,X)\mid A=a',X,S=s\}\mid S=s\bigr]
\end{align*}
We estimated $\widehat{\psi}_s(a,a')$ (and consequently $\widehat{\text{NDE}}_s$ and $\widehat{\text{NIE}}_s$) using the influence-function-based estimator proposed by \citet{ghassami2025causal}. This estimator accommodates flexible, nonparametric machine learning methods for nuisance function estimation while still yielding valid statistical inference under appropriate regularity conditions. Cross-fitting is used to estimate all required nuisance functions, including the propensity score $p_a(x)= P(A = a \mid X = x)$, bridge functions $q_a(z,x)$ and $h_a(w,x)$, and conditional expectation functions $\mu_a(x)=\mathbb{E}[Y \mid A=a,X=x]$ and $\eta_{a',a}(x)=\mathbb{E}[h_{a'}(W,X) \mid A = a, X= x]$ \citep{chernozhukov2018double}. Although we estimated $p_a(x)$, the known randomization probabilities could alternatively be used. $p_a(x)$, $\mu_a(x)$, and $\eta_{a',a}(x)$ were estimated using SuperLearner with cross-validation and a library consisting of the sample mean, generalized linear models, and generalized additive models. In particular, the estimator for $\eta_{a',a}(x)$ used predicted values from the estimated bridge function $\widehat{h}_{a'}(W,X)$. The bridge functions $q_a(z,x)$ and $h_{a'}(w,x)$ were estimated nonparametrically in a reproducing kernel Hilbert space (RKHS) with Gaussian kernels using an adversarial approach \citep{ghassami2025causal, olivas2025source}, with tuning parameters selected by cross-validation. When $a=a'$, the estimands reduce to standard counterfactual mean outcomes and bridge function estimation is not required. Additional estimation details are provided in Appendix \ref{app:C-effects}. We report effect estimates on the risk difference scale together with 95\% confidence intervals based on the empirical variance of the estimated influence function.

As previously discussed, identification of natural direct and indirect effects relies on a controversial cross-world independence assumption, which generally fails in the presence of treatment-induced mediator-outcome confounding \citep{vanderweele2014effect}. In our setting, early pregnancy complications (other than injection site reactions) may represent one such treatment-induced confounder. RIDEs and RIIEs provide an alternative notion of a mediation effect that does not require this assumption. We establish novel identification results for these effects in our hidden mediation setting, both without treatment-induced mediator-outcome confounding (Appendix \ref{app:B-interventional}) and with measured treatment-induced mediator-outcome confounding (Appendix \ref{app:B-interventional-L}). In the former setting, the identifying statistical functionals coincide with those for the NDE and NIE. Thus, even without imposing the cross-world independence assumption, the estimates of the NDE and NIE can still be interpreted as estimates of the RIDE and RIIE with no treatment-induced mediator-outcome confounding. Estimation of randomized interventional effects in our setting under measured treatment-induced mediator-outcome confounding is discussed in Appendix \ref{app:C-interventional-L}.

Another alternative ``direct effect" estimand that does not rely on the cross-world independence assumption is the CDE. Here, we considered the CDE of 17OHP-C under a hypothetical intervention eliminating injection site reactions within each trial,
\begin{align}
\label{eq:cde}
    CDE_s = \theta_{s}^Z(1,0)-\theta_{s}^Z(0,0),
\end{align}
where $\theta_{s}^Z(a,z) = \mathbb{E}[Y^Z(a, z) \mid S = s]$ denotes the mean potential outcome in trial $s$ under an intervention setting treatment assignment to $a$ and injection site reactions to $z$.

Although the CDE under a hypothetical intervention directly eliminating unblinding would be of greater interest, this estimand is not identifiable without additional assumptions because unblinding was unmeasured (Appendix \ref{app:B-cde-nonidentification}). However, the CDE in Equation \ref{eq:cde} is identified in our setup under standard assumptions (Appendix \ref{app:B-cde-identification}) and can be estimated using a standard influence-function-based estimator with flexible cross-fitted nuisance functions (Appendix \ref{app:C-cde-estimation}) \citep{robins1992identifiability, pearl2001direct, van2003unified}. In fact, under the additional assumptions that injection site reactions affect the outcome only through unblinding and that unblinding cannot occur in the absence of injection site reactions, the CDE in Equation \ref{eq:cde} can be interpreted as the CDE under a hypothetical intervention eliminating unblinding (Appendix \ref{app:B-cde-equivalence}). Comparing the CDEs across trials therefore provides a complementary notion of reconciliation to the one based on the NDEs in Equation \ref{eq:null}.

To assess reconciliation, we tested the null hypothesis
\begin{align}
\label{eq:null}
    H_0:\text{NDE}_1 - \text{NDE}_0=0
\end{align}
by constructing the estimate $\widehat{\text{NDE}}_1 - \widehat{\text{NDE}}_0$ and its corresponding 95\% CI using the methods described above. Under Equation \ref{eq:null}, the trials are reconcilable if the treatment effect not mediated through unblinding is the same in both trials, even if the indirect effect operating through unblinding, and consequently the total treatment effect, differ between the two trials. For the CDE, we similarly tested the null hypothesis
\begin{align}
\label{eq:nullCDE}
    H_0:\text{CDE}_1 - \text{CDE}_0=0
\end{align}
using the estimate $\widehat{\text{CDE}}_1 - \widehat{\text{CDE}}_0$ and its corresponding 95\% CI as described above.

We conducted a sensitivity analysis and subgroup analyses by race and BMI. First, we repeated the mediation analysis using gestational age at delivery (weeks) as a continuous outcome. Compared with the binary indicator of delivery before 37 weeks’ gestation, a continuous outcome may capture additional clinically relevant variation in treatment effects.

Second, we repeated the mediation analysis within prespecified subgroups, including non-Hispanic Black participants and participants with elevated prepregnancy body mass index (BMI $\geq 25~\mathrm{kg}/\mathrm{m}^2$). These analyses were motivated by prior evidence that the risk of preterm birth and the effectiveness of 17OHP-C may vary across subgroups, raising the possibility that mediation through unblinding could also differ across subgroups.

All analyses were conducted in R version 4.5.2. Code implementing the proximal hidden mediation estimators is available at https://github.com/danielxu19/Proximal-Mediation-Reconciliation. This secondary analysis of de-identified trial data was determined to be exempt by the Institutional Review Board at the University of Pennsylvania.

\section{Results}
\label{s:results}

Table \ref{tab:table1} summarizes baseline characteristics and proxy variables for each trial. Injection site pain (Meis: 151/443 [34.1\%] vs. PROLONG: 70/1,676 [4.2\%]) and other injection site reactions (Meis: 92/443 [20.8\%] vs. PROLONG: 106/1,676 [6.3\%]) were more common in Meis than in PROLONG, whereas prescription of any tocolytic, corticosteroid, or antibiotic was more common in PROLONG than in Meis (Meis: 65/443 [14.7\%] vs. PROLONG: 289/1,676 [17.2\%]).

Results from the primary analysis are presented in Table \ref{tab:results:main}. The estimated average treatment effects were $-18.3$\% (95\% CI: $-27.9$\%,$-8.6$\%) in Meis and 1.1\% (95\% CI: $-3.0$\%,5.3\%) in PROLONG, reproducing the previously reported finding of a statistically significant protective effect of 17OHP-C in Meis but little evidence of benefit in PROLONG.

Within Meis, the estimated NDE was $-19.2$\% (95\% CI: $-27.5$\%,$-10.8$\%), whereas the estimated NIE was 0.9\% (95\% CI: $-2.5$\%,4.3\%). These estimates suggest that little, if any, of the observed protective effect of 17OHP-C was mediated through unmeasured unblinding. Specifically, the estimated NDE indicates a 19.2\% reduction in preterm birth risk with 17OHP-C versus placebo if each participant’s unblinding status were set to its natural value under placebo. In contrast, the estimated NIE indicates that, with treatment fixed at 17OHP-C, setting each participant’s unblinding status to its natural value under 17OHP-C rather than placebo would increase risk by 0.9\%. Within PROLONG, the estimated NDE was $0.1$\% (95\% CI: $-3.6$\%,$3.8$\%), whereas the estimated NIE was 1.0\% (95\% CI: $-0.7$\%,2.7\%), suggesting very little evidence of either a direct effect of 17OHP-C or an indirect effect through unmeasured unblinding. As described in Appendix \ref{app:B-interventional}, these estimates may also be interpreted as randomized interventional direct and indirect effects, without requiring the cross-world independence assumption.

The estimated difference in NDEs between Meis and PROLONG was $-19.3$\% (95\% CI: $-28.4$\%,$-10.1$\%), providing strong evidence against the reconciliation hypothesis in Equation \ref{eq:null}. Under the proposed assumptions, the conflicting treatment effects cannot be reconciled under the hypothesis that the NDE of 17OHP-C is equal across trials.

Estimates of the CDE of 17OHP-C under an intervention eliminating injection site reactions are also presented in Table \ref{tab:results:main}. The estimated CDE was $-18.0$\% (95\% CI: $-31.2$\%,$-4.8$\%) in Meis and $-0.3$\% (95\% CI: $-4.7$\%,$4.1$\%) in PROLONG. In Meis, this estimate suggests that even if injection site reactions were eliminated for all participants, assignment to 17OHP-C rather than placebo would still reduce the expected risk of preterm birth by 18\%. In contrast, the estimated CDE in PROLONG indicates little evidence of a treatment effect. The estimated difference in CDEs between the two trials was $-17.7$\% (95\% CI: $-31.6$\%,$-3.7$\%), providing evidence against the reconciliation null hypothesis in Equation \ref{eq:nullCDE} that the CDEs are equal across trials.

Results from the sensitivity analysis using gestational age at delivery (weeks) as a continuous outcome are presented in Table \ref{tab:results:contoutcome}, where positive estimates indicate a protective effect of 17OHP-C (i.e., a longer gestational age at delivery). In Meis, the NDE was 0.76 weeks (95\% CI: $-0.08$,$1.60$) compared to the NIE of 0.25 weeks (95\% CI: $-0.20$,$0.69$). In PROLONG, the NDE was $-0.01$ weeks (95\% CI: $-0.32$,$0.30$) compared to the NIE of 0.13 weeks (95\% CI: $-0.03$,$0.28$). The estimated difference in NDEs between the two trials was 0.77 weeks (95\% CI: $-0.12$, 1.67). Although this confidence interval includes the null and may indicate a greater degree of reconciliation, such a result should be interpreted cautiously. In general, failure to reject the null hypothesis may simply reflect limited statistical power, for example due to weak or uninformative proxy variables \citep{altman1995statistics, imai2008misunderstandings, wasserstein2016asa}. Formal evidence supporting reconciliation would instead require an equivalence testing framework \citep{schuirmann1987comparison, westlake1972use}. We refer the reader to \citet{xu2026regression} for additional details.

Table \ref{tab:results:contoutcome} also presents estimates of the CDEs for the continuous outcome of gestational age at delivery. The estimated CDE was $0.29$ weeks (95\% CI: $-1.04$,$1.62$) in Meis and $0.19$ weeks (95\% CI: $-0.18$,$0.56$) in PROLONG, yielding a between trial difference of $0.09$ weeks (95\% CI: $-1.29$, 1.47). Although these findings suggest closer agreement between the trials' CDE estimates than between their ATE estimates, we again caution against claiming reconciliation without a formal equivalence testing framework.

Finally, subgroup analyses by race and prepregnancy BMI are presented in Tables \ref{tab:results:race} and \ref{tab:results:bmi}, respectively. As the results were broadly consistent with the primary analysis, we briefly highlight some key findings. Among non-Hispanic Black participants, the estimated differences in the NDE and CDE between the two trials were negative, although their 95\% CIs included the null. Among non-Hispanic White participants, both differences were also negative, but only the 95\% CI for the NDE difference excluded the null. Among participants with BMI $<25 \text{ kg/m}^2$, both differences were again negative, with 95\% CIs excluding the null. Across these subgroups, the NDE and CDE estimates suggested a greater ``direct" benefit of 17OHP-C in Meis than in PROLONG, with nearly all direct effect estimates in Meis being statistically significant. In contrast, among participants with BMI $\geq25 \text{ kg/m}^2$, the estimated ATE, NDE, and CDE in Meis were closer to the null, and all corresponding 95\% CIs included the null. The estimated differences in these effects between trials were likewise closer to the null, with 95\% CIs that included the null.

\section{Discussion}
\label{s:discussion}

In this study, we investigated whether differential unmeasured unblinding could explain the conflicting findings of the Meis and PROLONG trials. Our primary results suggested that nearly all of the observed protective treatment effect in Meis operated through pathways other than unblinding. Moreover, the estimated NIE in Meis could not account for the large difference in total treatment effects between the two trials. Together with other secondary and sensitivity analyses, our results suggest that differential unmeasured unblinding, as captured by the proposed proxy structure, is unlikely to fully reconcile the two trials, although analyses of gestational age on the continuous scale and among participants with BMI $\geq25 \text{ kg/m}^2$ suggest the greatest potential for reconciliation of the direct effect of 17OHP-C.

A primary limitation concerns the assumptions required for identification. Our proximal mediation framework generally requires that treatment assignment and injection site reactions affect the medication prescription proxy only through unblinding, and that injection site reactions affect preterm birth only through unblinding. The former assumption could be violated if treatment assignment affected the occurrence of pregnancy complications for which these medications are prescribed. The latter could be violated if injection site reactions led participants to delay or discontinue injections because of discomfort, regardless of unblinding. Broadening the latent mediator beyond unblinding could accommodate some of these pathways, but would change the scientific question and require that the measured proxies adequately capture the broader mediator. Developing sensitivity analyses to assess the impact of violations of these assumptions is an important direction for future research.

Differences in measurement and reporting across the two trials further complicate interpretation. Meis was conducted exclusively in the United States, whereas PROLONG enrolled participants across nine countries. Differences in adverse event ascertainment, participant symptom reporting, and clinician documentation could contribute to systematic differences in the observed proxy variables, particularly injection site reactions. Changes in clinical practice over time could also lead to differences in the medication prescription proxy between trials. Future work should assess the robustness of our findings to these differences.

Another important limitation is that our primary reconciliation analysis compares marginal NDEs between two trial populations with substantially different baseline covariate distributions. If baseline covariates modify the NDE, then marginal NDEs may differ even when conditional NDEs are identical across trials. We therefore extend our framework to standardize natural direct and indirect effects to a common covariate distribution in Appendices \ref{app:B-standardized} and \ref{app:C-standardized}. This extension relates to methods for transported mediation effects \citep{rudolph2021transporting, rudolph2022efficiently} and builds on previous work to reconcile the Meis and PROLONG trials by accounting for differences in effect modifiers \citep{virkud2025clarifying, xu2026regression}. The standardized analyses (\ref{app:A-additional}) yield conclusions broadly consistent with those in the main text.

Finally, differential unblinding is only one of several plausible explanations for the conflicting trial results \citep{virkud2025clarifying}. Conflicting randomized trials are likely to arise from multiple mechanisms operating simultaneously \citep{jin2026diagnosing}. Developing reconciliation frameworks that can jointly accommodate multiple explanations would help clinicians and regulators better evaluate conflicting evidence from trials.

\bibliography{sample}

\clearpage

\begin{table}[!h]
\centering
\caption{\label{tab:table1}Distribution of baseline covariates and proxies by assigned treatment and trial. ``Other" injection site reactions are defined in Appendix \ref{app:A-measurement}. Any of the above medications indicates prescription of at least one of corticosteroids, antibiotics, or tocolytics. Categorical variables are reported as n (percent).}
\centering
\resizebox{\ifdim\width>\linewidth\linewidth\else\width\fi}{!}{
\begin{tabular}[t]{>{\raggedright\arraybackslash}p{16em}>{\raggedright\arraybackslash}p{6em}>{\raggedright\arraybackslash}p{6em}>{\raggedright\arraybackslash}p{6em}>{\raggedright\arraybackslash}p{6em}}
\toprule
\multicolumn{1}{c}{ } & \multicolumn{2}{c}{PROLONG} & \multicolumn{2}{c}{Meis} \\
\cmidrule(l{3pt}r{3pt}){2-3} \cmidrule(l{3pt}r{3pt}){4-5}
 & Placebo (N=568) & 17OHP-C (N=1108) & Placebo (N=149) & 17OHP-C (N=294)\\
\midrule
\addlinespace[0.3em]
\multicolumn{5}{l}{\textbf{Baseline covariates}}\\
Age (years), mean (SD) & 29.9 (5.23) & 29.9 (5.15) & 26.5 (5.42) & 26.1 (5.70)\\
Race &  &  &  & \\
\hspace{1em}Non-Hispanic Black & 40 (7.0\%) & 70 (6.3\%) & 89 (59.7\%) & 177 (60.2\%)\\
\hspace{1em}Non-Hispanic White & 445 (78.3\%) & 895 (80.8\%) & 34 (22.8\%) & 74 (25.2\%)\\
\hspace{1em}Hispanic, Asian, or Other & 83 (14.6\%) & 143 (12.9\%) & 26 (17.4\%) & 43 (14.6\%)\\
Gestational age at qualifying prior PTB in weeks, mean (SD) & 31.6 (4.13) & 31.3 (4.32) & 31.3 (4.21) & 30.5 (4.62)\\
Number of prior PTBs, mean (SD) & 1.14 (0.407) & 1.18 (0.543) & 1.61 (0.913) & 1.40 (0.726)\\
Married or living with partner & 512 (90.1\%) & 993 (89.6\%) & 69 (46.3\%) & 149 (50.7\%)\\
Prepregnancy BMI, mean (SD) & 24.7 (8.70) & 24.3 (7.09) & 26.0 (6.96) & 27.0 (7.94)\\
Years of education, mean (SD) & 13.0 (2.37) & 13.1 (2.37) & 12.0 (2.34) & 11.8 (2.25)\\
Smoking during pregnancy & 40 (7.0\%) & 90 (8.1\%) & 30 (20.1\%) & 65 (22.1\%)\\
Alcohol use during pregnancy & 17 (3.0\%) & 23 (2.1\%) & 10 (6.7\%) & 26 (8.8\%)\\
Substance use during pregnancy & 8 (1.4\%) & 14 (1.3\%) & 4 (2.7\%) & 11 (3.7\%)\\
\addlinespace[0.3em]
\hline
\multicolumn{5}{l}{\textbf{Proxies}}\\
\textbf{Injection site reactions} & \textbf{} & \textbf{} & \textbf{} & \textbf{}\\
\hspace{1em}Pain & 29 (5.1\%) & 41 (3.7\%) & 50 (33.6\%) & 101 (34.4\%)\\
\hspace{1em}Other & 32 (5.6\%) & 74 (6.7\%) & 23 (15.4\%) & 69 (23.5\%)\\
\textbf{Prescription of medications for preterm birth} & \textbf{} & \textbf{} & \textbf{} & \textbf{}\\
\hspace{1em}Corticosteroids & 51 (9.0\%) & 99 (8.9\%) & 12 (8.1\%) & 21 (7.1\%)\\
\hspace{1em}Antibiotics & 3 (0.5\%) & 17 (1.5\%) & 6 (4.0\%) & 12 (4.1\%)\\
\hspace{1em}Tocolytics & 68 (12.0\%) & 151 (13.6\%) & 18 (12.1\%) & 38 (12.9\%)\\
\hspace{1em}Any of the above & 92 (16.2\%) & 197 (17.8\%) & 20 (13.4\%) & 45 (15.3\%)\\
\bottomrule
\end{tabular}}
\end{table}

\begin{table}[!h]
\centering
\caption{\label{tab:results:main}Coefficient estimates with 95\% confidence intervals (CIs) for the primary analysis. CIs are constructed using standard errors based on the empirical variance of the estimated influence functions.}
\centering
\resizebox{\ifdim\width>\linewidth\linewidth\else\width\fi}{!}{
\fontsize{10}{12}\selectfont
\begin{tabular}[t]{llll}
\toprule
 & Meis & PROLONG & Difference\\
\midrule
$\psi(1,1)$ & 0.363 (0.307, 0.418) & 0.230 (0.205, 0.255) & 0.133 (0.072, 0.193)\\
$\psi(0,0)$ & 0.545 (0.466, 0.625) & 0.219 (0.185, 0.253) & 0.326 (0.240, 0.413)\\
$\psi(1,0)$ & 0.354 (0.323, 0.385) & 0.220 (0.200, 0.239) & 0.134 (0.097, 0.170)\\
ATE & -0.183 (-0.279, -0.086) & 0.011 (-0.030, 0.053) & -0.194 (-0.299, -0.089)\\
NDE & -0.192 (-0.275, -0.108) & 0.001 (-0.036, 0.038) & -0.193 (-0.284, -0.101)\\
NIE & 0.009 (-0.025, 0.043) & 0.010 (-0.007, 0.027) & -0.001 (-0.039, 0.036)\\
\midrule
$\theta^Z(1,0)$ & 0.361 (0.286, 0.435) & 0.225 (0.199, 0.251) & 0.136 (0.057, 0.215)\\
$\theta^Z(0,0)$ & 0.541 (0.431, 0.650) & 0.228 (0.192, 0.264) & 0.313 (0.198, 0.427)\\
CDE & -0.180 (-0.312, -0.048) & -0.003 (-0.047, 0.041) & -0.177 (-0.316, -0.037)\\
\bottomrule
\end{tabular}}
\end{table}

\begin{table}[!h]
\centering
\caption{\label{tab:results:contoutcome}Coefficient estimates with 95\% confidence intervals (CIs) for the secondary analysis where the outcome is gestational age at delivery in weeks. CIs are constructed using standard errors based on the empirical variance of the estimated influence functions.}
\centering
\resizebox{\ifdim\width>\linewidth\linewidth\else\width\fi}{!}{
\fontsize{10}{12}\selectfont
\begin{tabular}[t]{llll}
\toprule
 & Meis & PROLONG & Difference\\
\midrule
$\psi(1,1)$ & 36.193 (35.639, 36.748) & 37.729 (37.548, 37.910) & -1.535 (-2.119, -0.952)\\
$\psi(0,0)$ & 35.186 (34.445, 35.927) & 37.618 (37.335, 37.901) & -2.432 (-3.226, -1.639)\\
$\psi(1,0)$ & 35.945 (35.534, 36.355) & 37.604 (37.448, 37.760) & -1.660 (-2.099, -1.221)\\
ATE & 1.008 (0.084, 1.931) & 0.111 (-0.224, 0.445) & 0.897 (-0.086, 1.879)\\
NDE & 0.759 (-0.081, 1.599) & -0.014 (-0.324, 0.296) & 0.773 (-0.122, 1.668)\\
NIE & 0.249 (-0.196, 0.694) & 0.125 (-0.032, 0.282) & 0.124 (-0.347, 0.596)\\
\midrule
$\theta^Z(1,0)$ & 35.492 (34.605, 36.378) & 37.722 (37.526, 37.918) & -2.230 (-3.138, -1.322)\\
$\theta^Z(0,0)$ & 35.205 (34.216, 36.194) & 37.528 (37.213, 37.842) & -2.323 (-3.361, -1.286)\\
CDE & 0.287 (-1.042, 1.617) & 0.194 (-0.175, 0.563) & 0.093 (-1.287, 1.473)\\
\bottomrule
\end{tabular}}
\end{table}

\begin{table}[!h]
\centering
\caption{\label{tab:results:race}Coefficient estimates with 95\% confidence intervals (CIs) for subgroup analyses stratified by non-Hispanic Black and non-Hispanic White participants. For the natural effect analyses, the three medication proxies were replaced by a single indicator of any tocolytic, corticosteroid, or antibiotic prescription to address positivity issues. Additionally, due to positivity issues, substance use was omitted from baseline covariates in all analyses, and alcohol use was omitted for the analysis considering non-Hispanic Black participants in PROLONG.}
\centering
\resizebox{\ifdim\width>\linewidth\linewidth\else\width\fi}{!}{
\fontsize{10}{12}\selectfont
\begin{tabular}[t]{>{\raggedright\arraybackslash}p{12em}lll}
\toprule
 & Meis & PROLONG & Difference\\
\midrule
\addlinespace[0.3em]
\multicolumn{4}{l}{\textbf{Non-Hispanic Black}}\\
\hspace{1em}$\psi(1,1)$ & 0.359 (0.287, 0.430) & 0.369 (0.254, 0.485) & -0.011 (-0.147, 0.125)\\
\hspace{1em}$\psi(0,0)$ & 0.486 (0.385, 0.586) & 0.323 (0.166, 0.480) & 0.162 (-0.025, 0.349)\\
\hspace{1em}$\psi(1,0)$ & 0.335 (0.299, 0.371) & 0.331 (0.274, 0.387) & 0.004 (-0.063, 0.071)\\
\hspace{1em}ATE & -0.127 (-0.250, -0.004) & 0.046 (-0.150, 0.242) & -0.173 (-0.404, 0.058)\\
\hspace{1em}NDE & -0.151 (-0.257, -0.044) & 0.008 (-0.137, 0.152) & -0.158 (-0.338, 0.021)\\
\hspace{1em}NIE & 0.024 (-0.021, 0.069) & 0.038 (-0.068, 0.145) & -0.015 (-0.130, 0.101)\\
\cmidrule(l{\dimexpr\tabcolsep+1em\relax}){1-4}
\hspace{1em}$\theta^Z(1,0)$ & 0.351 (0.261, 0.441) & 0.392 (0.240, 0.544) & -0.041 (-0.218, 0.135)\\
\hspace{1em}$\theta^Z(0,0)$ & 0.537 (0.406, 0.668) & 0.456 (0.228, 0.683) & 0.081 (-0.182, 0.344)\\
\hspace{1em}CDE & -0.186 (-0.344, -0.028) & -0.063 (-0.340, 0.213) & -0.123 (-0.441, 0.195)\\
\midrule
\addlinespace[0.3em]
\multicolumn{4}{l}{\textbf{Non-Hispanic White}}\\
\hspace{1em}$\psi(1,1)$ & 0.384 (0.270, 0.498) & 0.213 (0.186, 0.240) & 0.171 (0.054, 0.288)\\
\hspace{1em}$\psi(0,0)$ & 0.611 (0.423, 0.800) & 0.201 (0.163, 0.238) & 0.411 (0.218, 0.603)\\
\hspace{1em}$\psi(1,0)$ & 0.350 (0.298, 0.401) & 0.194 (0.174, 0.215) & 0.155 (0.100, 0.210)\\
\hspace{1em}ATE & -0.228 (-0.448, -0.007) & 0.012 (-0.034, 0.058) & -0.240 (-0.465, -0.015)\\
\hspace{1em}NDE & -0.262 (-0.444, -0.080) & -0.006 (-0.048, 0.035) & -0.256 (-0.442, -0.069)\\
\hspace{1em}NIE & 0.034 (-0.072, 0.141) & 0.018 (0.001, 0.036) & 0.016 (-0.092, 0.124)\\
\cmidrule(l{\dimexpr\tabcolsep+1em\relax}){1-4}
\hspace{1em}$\theta^Z(1,0)$ & 0.372 (0.203, 0.542) & 0.213 (0.185, 0.241) & 0.159 (-0.013, 0.331)\\
\hspace{1em}$\theta^Z(0,0)$ & 0.611 (0.265, 0.956) & 0.199 (0.161, 0.238) & 0.411 (0.064, 0.759)\\
\hspace{1em}CDE & -0.238 (-0.623, 0.146) & 0.014 (-0.034, 0.061) & -0.252 (-0.639, 0.135)\\
\bottomrule
\end{tabular}}
\end{table}

\begin{table}[!h]
\centering
\caption{\label{tab:results:bmi}Coefficient estimates with 95\% confidence intervals (CIs) for subgroup analyses stratified by pre-pregnancy BMI $\geq25$ versus $<25\,\mathrm{kg/m^2}$. For the natural effect analyses, the three medication proxies were replaced by a single indicator of any tocolytic, corticosteroid, or antibiotic prescription to address positivity issues. Additionally, due to positivity issues, alcohol use and substance use were omitted from baseline covariates for the analyses considering the BMI $\geq25$ subgroup, and substance use was omitted from baseline covariates for the analysis considering BMI $<25$ participants in Meis.}
\centering
\resizebox{\ifdim\width>\linewidth\linewidth\else\width\fi}{!}{
\fontsize{10}{12}\selectfont
\begin{tabular}[t]{llll}
\toprule
 & Meis & PROLONG & Difference\\
\midrule
\addlinespace[0.3em]
\multicolumn{4}{l}{\textbf{BMI $\geq25$}}\\
\hspace{1em}$\psi(1,1)$ & 0.377 (0.299, 0.454) & 0.262 (0.219, 0.306) & 0.114 (0.025, 0.203)\\
\hspace{1em}$\psi(0,0)$ & 0.433 (0.308, 0.559) & 0.282 (0.219, 0.344) & 0.152 (0.011, 0.292)\\
\hspace{1em}$\psi(1,0)$ & 0.358 (0.325, 0.391) & 0.262 (0.229, 0.295) & 0.096 (0.050, 0.143)\\
\hspace{1em}ATE & -0.057 (-0.204, 0.091) & -0.019 (-0.096, 0.057) & -0.037 (-0.204, 0.129)\\
\hspace{1em}NDE & -0.075 (-0.203, 0.053) & -0.020 (-0.087, 0.048) & -0.056 (-0.200, 0.089)\\
\hspace{1em}NIE & 0.019 (-0.036, 0.074) & 0.001 (-0.032, 0.033) & 0.018 (-0.045, 0.082)\\
\cmidrule(l{\dimexpr\tabcolsep+1em\relax}){1-4}
\hspace{1em}$\theta^Z(1,0)$ & 0.415 (0.313, 0.518) & 0.256 (0.209, 0.302) & 0.160 (0.047, 0.272)\\
\hspace{1em}$\theta^Z(0,0)$ & 0.441 (0.287, 0.595) & 0.317 (0.242, 0.393) & 0.124 (-0.048, 0.296)\\
\hspace{1em}CDE & -0.025 (-0.211, 0.160) & -0.062 (-0.151, 0.027) & 0.036 (-0.169, 0.242)\\
\midrule
\addlinespace[0.3em]
\multicolumn{4}{l}{\textbf{BMI $<25$}}\\
\hspace{1em}$\psi(1,1)$ & 0.354 (0.274, 0.434) & 0.211 (0.181, 0.240) & 0.143 (0.058, 0.229)\\
\hspace{1em}$\psi(0,0)$ & 0.639 (0.533, 0.744) & 0.186 (0.146, 0.227) & 0.453 (0.340, 0.565)\\
\hspace{1em}$\psi(1,0)$ & 0.322 (0.289, 0.354) & 0.196 (0.174, 0.218) & 0.126 (0.087, 0.165)\\
\hspace{1em}ATE & -0.285 (-0.417, -0.153) & 0.024 (-0.025, 0.074) & -0.309 (-0.450, -0.168)\\
\hspace{1em}NDE & -0.317 (-0.427, -0.208) & 0.009 (-0.034, 0.053) & -0.327 (-0.445, -0.209)\\
\hspace{1em}NIE & 0.032 (-0.023, 0.088) & 0.015 (-0.004, 0.034) & 0.018 (-0.041, 0.076)\\
\cmidrule(l{\dimexpr\tabcolsep+1em\relax}){1-4}
\hspace{1em}$\theta^Z(1,0)$ & 0.297 (0.190, 0.404) & 0.204 (0.173, 0.234) & 0.093 (-0.018, 0.204)\\
\hspace{1em}$\theta^Z(0,0)$ & 0.622 (0.463, 0.780) & 0.190 (0.149, 0.230) & 0.432 (0.269, 0.596)\\
\hspace{1em}CDE & -0.325 (-0.516, -0.134) & 0.014 (-0.036, 0.065) & -0.339 (-0.537, -0.142)\\
\bottomrule
\end{tabular}}
\end{table}

\clearpage

\appendix

\printappendixtoc

\appsection{Appendix A}
\label{app:appendixA}
\appsubsection{Measurement and assessment of proxies}
\label{app:A-measurement}

Injection site reactions were categorized based on the MedDRA preferred term for adverse events. Table \ref{tab:injection-site-reactions} lists all injection site reactions reported in either trial before participants with missing outcome or covariate data were excluded. We assessed reactions reported at any study visit and classified participants as having an ``other'' injection site reaction if they reported at least one reaction other than pain. Because most reactions were reported early in follow-up, restricting ascertainment to an earlier window would likely yield similar results and was not formally considered in our primary analysis.

Ascertainment of tocolytic, corticosteroid, and antibiotic prescriptions differed between Meis and PROLONG. In Meis, we identified tocolytic and corticosteroid prescriptions recorded at hospital admissions or labor and delivery visits between randomization and 37 weeks' gestation at which at least six contractions per hour were observed. We identified antibiotic prescriptions between randomization and delivery using reason codes for preterm birth, PROM, or chorioamnionitis. In PROLONG, we identified tocolytics from records classified as tocolytic medications. We identified corticosteroids from concomitant medication records for betamethasone or dexamethasone with an indication specifically related to preterm birth (e.g., preventing infant respiratory distress syndrome, promoting fetal lung maturation). Other indications (e.g., allergy-related) were excluded. We identified antibiotics from concomitant medication records classified as systemic antibacterials whose indication contained ``chorioamnionitis,'' ``preterm,'' ``pre-term,'' ``premature,'' ``amniotic,'' or ``PROM.''

\begin{table}[!h]
\centering
\caption{\label{tab:injection-site-reactions}Complete list of injection site reactions by treatment assignment and trial.}
\centering
\resizebox{\ifdim\width>\linewidth\linewidth\else\width\fi}{!}{
\begin{tabular}[t]{>{\raggedright\arraybackslash}p{16em}>{\raggedright\arraybackslash}p{6em}>{\raggedright\arraybackslash}p{6em}>{\raggedright\arraybackslash}p{6em}>{\raggedright\arraybackslash}p{6em}}
\toprule
\multicolumn{1}{c}{ } & \multicolumn{2}{c}{PROLONG} & \multicolumn{2}{c}{Meis} \\
\cmidrule(l{3pt}r{3pt}){2-3} \cmidrule(l{3pt}r{3pt}){4-5}
Reaction & Placebo\newline{}\mbox{(N = 578)} & 17OHP-C\newline{}\mbox{(N = 1130)} & Placebo\newline{}\mbox{(N = 153)} & 17OHP-C\newline{}\mbox{(N = 310)}\\
\midrule
anaesthesia & 0 (0.0\%) & 0 (0.0\%) & 0 (0.0\%) & 2 (0.6\%)\\
coldness & 0 (0.0\%) & 0 (0.0\%) & 0 (0.0\%) & 1 (0.3\%)\\
discomfort & 3 (0.5\%) & 10 (0.9\%) & 0 (0.0\%) & 0 (0.0\%)\\
erosion & 0 (0.0\%) & 1 (0.1\%) & 0 (0.0\%) & 0 (0.0\%)\\
erythema & 1 (0.2\%) & 10 (0.9\%) & 1 (0.7\%) & 3 (1.0\%)\\
haematoma & 5 (0.9\%) & 10 (0.9\%) & 0 (0.0\%) & 0 (0.0\%)\\
inflammation & 0 (0.0\%) & 2 (0.2\%) & 0 (0.0\%) & 1 (0.3\%)\\
irritation & 0 (0.0\%) & 1 (0.1\%) & 5 (3.3\%) & 4 (1.3\%)\\
nodule & 9 (1.6\%) & 19 (1.7\%) & 3 (2.0\%) & 14 (4.5\%)\\
pain & 29 (5.0\%) & 41 (3.6\%) & 50 (32.7\%) & 108 (34.8\%)\\
pruritus & 24 (4.2\%) & 42 (3.7\%) & 5 (3.3\%) & 18 (5.8\%)\\
rash & 2 (0.3\%) & 6 (0.5\%) & 0 (0.0\%) & 2 (0.6\%)\\
reaction & 1 (0.2\%) & 4 (0.4\%) & 1 (0.7\%) & 4 (1.3\%)\\
swelling & 2 (0.3\%) & 5 (0.4\%) & 12 (7.8\%) & 53 (17.1\%)\\
urticaria & 0 (0.0\%) & 0 (0.0\%) & 0 (0.0\%) & 2 (0.6\%)\\
warmth & 0 (0.0\%) & 2 (0.2\%) & 0 (0.0\%) & 1 (0.3\%)\\
\bottomrule
\end{tabular}}
\end{table}

\appsubsection{Additional data analyses}
\label{app:A-additional}

Table \ref{tab:results:alternative} presents results under the alternative natural effect decomposition involving $\psi(0,1)$ rather than $\psi(1,0)$, and Table \ref{tab:results:standardized} presents results applying the standardized natural effect estimators proposed in Appendix \ref{app:C-standardized}.

\begin{table}[!h]
\centering
\caption{\label{tab:results:alternative}Coefficient estimates with 95\% confidence intervals (CIs) for the analysis with a binary outcome under the alternative natural effect decomposition: $\mathrm{NDE}=\psi(1,1)-\psi(0,1)$ and $\mathrm{NIE}=\psi(0,1)-\psi(0,0)$. CIs are constructed using standard errors based on the empirical variance of the estimated influence functions.}
\centering
\resizebox{\ifdim\width>\linewidth\linewidth\else\width\fi}{!}{
\fontsize{10}{12}\selectfont
\begin{tabular}[t]{llll}
\toprule
 & Meis & PROLONG & Difference\\
\midrule
$\psi(1,1)$ & 0.363 (0.307, 0.418) & 0.230 (0.205, 0.255) & 0.133 (0.072, 0.193)\\
$\psi(0,0)$ & 0.545 (0.466, 0.625) & 0.219 (0.185, 0.253) & 0.326 (0.240, 0.413)\\
$\psi(0,1)$ & 0.539 (0.503, 0.576) & 0.215 (0.193, 0.236) & 0.324 (0.282, 0.367)\\
ATE & -0.183 (-0.279, -0.086) & 0.011 (-0.030, 0.053) & -0.194 (-0.299, -0.089)\\
NDE & -0.176 (-0.241, -0.112) & 0.015 (-0.016, 0.047) & -0.192 (-0.264, -0.120)\\
NIE & -0.006 (-0.064, 0.052) & -0.004 (-0.025, 0.017) & -0.002 (-0.064, 0.059)\\
\bottomrule
\end{tabular}}
\end{table}

\begin{table}[!h]
\centering
\caption{\label{tab:results:standardized}Coefficient estimates with 95\% confidence intervals (CIs) for the standardized natural effect analysis with a binary outcome. CIs are constructed using standard errors based on the empirical variance of the estimated influence functions.}
\fontsize{10}{12}\selectfont

\begin{tabular}[t]{llll}
\toprule
 & Meis & PROLONG $\to$ Meis & Difference\\
\midrule
$\psi(1,1)$ & 0.363 (0.307, 0.418) & 0.416 (0.298, 0.535) & -0.054 (-0.184, 0.076)\\
$\psi(0,0)$ & 0.545 (0.466, 0.625) & 0.344 (0.208, 0.481) & 0.201 (0.043, 0.359)\\
$\psi(1,0)$ & 0.354 (0.323, 0.385) & 0.350 (0.282, 0.418) & 0.004 (-0.070, 0.078)\\
ATE & -0.183 (-0.279, -0.086) & 0.072 (-0.108, 0.252) & -0.255 (-0.460, -0.050)\\
NDE & -0.192 (-0.275, -0.108) & 0.006 (-0.126, 0.137) & -0.197 (-0.354, -0.041)\\
NIE & 0.009 (-0.025, 0.043) & 0.067 (-0.060, 0.194) & -0.058 (-0.189, 0.074)\\
\midrule

 & PROLONG & Meis $\to$ PROLONG & Difference\\
\midrule
$\psi(1,1)$ & 0.230 (0.205, 0.255) & 0.359 (0.259, 0.459) & -0.129 (-0.232, -0.026)\\
$\psi(0,0)$ & 0.219 (0.185, 0.253) & 0.629 (0.487, 0.770) & -0.410 (-0.556, -0.264)\\
$\psi(1,0)$ & 0.220 (0.200, 0.239) & 0.357 (0.306, 0.408) & -0.137 (-0.192, -0.083)\\
ATE & 0.011 (-0.030, 0.053) & -0.270 (-0.443, -0.096) & 0.281 (0.102, 0.459)\\
NDE & 0.001 (-0.036, 0.038) & -0.272 (-0.423, -0.120) & 0.273 (0.117, 0.428)\\
NIE & 0.010 (-0.007, 0.027) & 0.002 (-0.072, 0.077) & 0.008 (-0.068, 0.085)\\
\bottomrule
\end{tabular}
\end{table}

\appsection{Appendix B}
\label{app:appendixB}

\appsubsection{Additional notation and estimands}
\label{app:B-notation}
We continue with the notation from the main text: $S$ indexes the trial ($1$: Meis; $0$: PROLONG), $A$ is treatment assignment ($0$: placebo; $1$: 17OHP-C), $X$ denotes baseline covariates, $M$ denotes unmeasured unblinding, $Z$ denotes injection site reactions, $W$ denotes the medication proxy, and $Y$ denotes the outcome. When $Z$ is a vector, $z=0$ denotes the vector with every component equal to zero. In the extension with intermediate confounding (Appendix Section \ref{app:B-interventional-L}), $L$ denotes an observed treatment-induced mediator-outcome confounder.

Let $Y(a)$, $M(a)$, and $Z(a)$ denote the potential outcome, unblinding status, and injection site reaction level under treatment assignment $a$, respectively. Let $Y(a,m)$ denote the potential outcome under joint intervention on treatment and unblinding, and $Y(a,z,m)$ the potential outcome under joint intervention on treatment, injection site reaction level, and unblinding status. Similarly, let $M(a,z)$ denote potential unblinding under assignment $a$ and injection site reaction level $z$. To distinguish interventions on injection site reactions from interventions on hidden unblinding, we introduce $Y^Z(a,z)$ for the former and $Y^M(a,m) := Y(a,m)$ for the latter. By composition,
\begin{align*}
Y(a)&=Y\{a,M(a)\},\\
Y^Z(a,z)&=Y\{a,z,M(a,z)\},\\
Y^M(a,m)&=Y\{a,Z(a),m\}.
\end{align*}

\appsubsection{Identification of the ATE, NDE, and NIE}
\label{app:B-natural-identification}
\begin{assumption}
\label{assump:ATEidentification}
For each $a,s \in \{0,1\}$,
\begin{enumerate}[(i)] 
    \item Consistency: $Y=Y(a)$ if $A=a$
    \item Conditional exchangeability: $Y(a)\indep A\mid X,S=s$
    \item Positivity: $0<\Pr(A=1\mid X,S=s)<1$ almost surely
\end{enumerate}
\end{assumption}
Under Assumption \ref{assump:ATEidentification}, the following standard identification result for the ATE holds:
\[
\mu_s(a):=\mathbb{E}\{Y(a)\mid S=s\}
=\mathbb{E}\bigl\{\mathbb{E}(Y\mid A=a,X,S=s)\mid S=s\bigr\},
\qquad
\mathrm{ATE}_s=\mu_s(1)-\mu_s(0).
\]

\paragraph{Additional assumptions for the NDE and NIE.}
We adapt the mediation model of \citet{ghassami2025causal}.
\begin{assumption}
\label{assump:NDENIEidentification}
For each $a,a',s \in \{0,1\}$,
\begin{enumerate}[(i)]
    \item Consistency: $M=M(a)$ if $A=a$ and 
$Y=Y(a,m)$ if $(A,M)=(a,m)$
    \item Conditional exchangeability. For $a \in \{0,1\}$,
    \begin{align*}
        &\{Y(a,m), M(a)\} \indep A\mid X,S=s,\\
        &Y(a,m) \indep M \mid A=a, X, S=s,\\
        &Y(a,m)\indep M(a')\mid X,S=s
    \end{align*}
    \item Positivity: $0<\mathrm{Pr}(A=1\mid M,X,S=s)<1$ almost surely
\end{enumerate}
\end{assumption}
Because $M$ is unobserved, we additionally impose the following assumptions regarding the proxies, completeness, and the existence of a bridge function from \citet{ghassami2025causal}.
\begin{assumption}
\label{assump:NDENIEproxycompletenessbridge}
Assume
\begin{enumerate}[(i)]
    \item For each $s \in \{0,1\}$,
    \begin{align*}
        Y\indep Z\mid A,M,X,S=s, \qquad  W\indep (A,Z)\mid M,X,S=s
    \end{align*}
    \item For each $a,x,s$ and every square-integrable $g$,
\[
\mathbb{E}\{g(M)\mid Z,A=a,X=x,S=s\}=0
\ \Longrightarrow\ g(M)=0\quad\text{almost surely},
\]
    \item There exists an integrable function $h_{s,a}(W,X)$ satisfying
\[
\mathbb{E}(Y\mid Z,A=a,X,S=s)
=\mathbb{E}\{h_{s,a}(W,X)\mid Z,A=a,X,S=s\}.
\]
\end{enumerate}
\end{assumption}
Under Assumptions \ref{assump:NDENIEidentification} and \ref{assump:NDENIEproxycompletenessbridge}, Theorem 1 from \citet{ghassami2025causal} identifies
\[
\psi_s(a,a')
=\mathbb{E}\bigl[\mathbb{E}\{h_{s,a}(W,X)\mid A=a',X,S=s\}\mid S=s\bigr].
\]
Consequently,
\[
\mathrm{NDE}_s=\psi_s(1,0)-\mu_s(0),\qquad
\mathrm{NIE}_s=\mu_s(1)-\psi_s(1,0).
\]

\appsubsection{Identification of the CDE under elimination of injection site reactions}
\label{app:B-cde-identification}
In this section, we provide identification results for
\[
\theta^Z_{s}(a,0)=\mathbb{E}\{Y^Z(a,0)\mid S=s\},\qquad
\mathrm{CDE}^Z_s=\theta^Z_{s}(1,0)-\theta^Z_{s}(0,0).
\]
Here $\mathrm{CDE}^Z_s$ is equivalent to $\mathrm{CDE}_s$ in the main text.
\begin{assumption}
\label{assump:CDEidentification}
\citep{robins1992identifiability,pearl2001direct,goetgeluk2008estimation}:
\begin{enumerate}
    \item Consistency: For $a \in \{0,1\}$, $Y=Y^Z(a,0)$ if $A=a$ and $Z=0$.
    \item Exchangeability: For $a \in \{0,1\}$, 
    \begin{align*}
        Y^Z(a,0)\indep A\mid X,S=s, \qquad Y^Z(a,0) \indep Z\mid A=a,X,S=s
    \end{align*}
    \item Positivity: For each $a,s\in\{0,1\}$,
    \begin{align*}
    &\Pr(A=a, Z = 0 \mid X, S=s)>0 \text{ almost surely},
    \end{align*}
\end{enumerate}
\end{assumption}
Let $m_{s,a}(x)=\mathbb{E}(Y\mid A=a,Z=0,X=x,S=s)$. Under Assumption \ref{assump:CDEidentification},
\begin{equation*}
\theta^Z_{s}(a,0)=\mathbb{E}\{m_{s,a}(X)\mid S=s\},\qquad
\mathrm{CDE}^Z_s=\mathbb{E}\{m_{s,1}(X)-m_{s,0}(X)\mid S=s\}.
\end{equation*}
\begin{proof}
By iterated expectation, exchangeability, and consistency,
\begin{align*}
\theta^Z_{s}(a,0)
&=\mathbb{E}\bigl[\mathbb{E}\{Y^Z(a,0)\mid X,S=s\}\mid S=s\bigr]\\
&=\mathbb{E}\bigl[\mathbb{E}\{Y^Z(a,0)\mid A=a,Z=0,X,S=s\}\mid S=s\bigr]\\
&=\mathbb{E}\{m_{s,a}(X)\mid S=s\}.
\end{align*}
\end{proof}

\appsubsection{When eliminating reactions is equivalent to eliminating unblinding}
\label{app:B-cde-equivalence}

The CDE under elimination of unblinding is
\[
\mathrm{CDE}^M_s=\mathbb{E}\{Y^M(1,0)-Y^M(0,0)\mid S=s\}.
\]
\begin{assumption}
    \label{assump:CDEequal}
    Assume
    \begin{enumerate}
    \item Exclusion restriction: For all relevant $a,z,z',m$,
    \[
    Y(a,z,m)=Y(a,z',m)\quad\text{almost surely}.
    \]
    \item Unblinding cannot occur without injection site reactions: For $a,s \in \{0,1\}$,
    \begin{equation}
    \Pr\{M(a,0)=0\mid S=s\}=1.
    \end{equation}
\end{enumerate}
\end{assumption}
The exclusion restriction states that injection site reactions affect the outcome only through unblinding.
\begin{theorem}
    \label{thm:CDEequal}
    Under Assumption \ref{assump:CDEequal}, $\mathrm{CDE}^Z_s=\mathrm{CDE}^M_s$.
\end{theorem}
\begin{proof}
    For each $a,s \in \{0,1\}$,
    \begin{align*}
\mathbb{E}[Y^Z(a,0) \mid S = s]
&=\mathbb{E}[Y\{a,0,M(a,0)\} \mid S = s]&&\text{(composition)}\\
&=\mathbb{E}[Y(a,0,0) \mid S = s]&&\text{(no unblinding at $z=0$)}\\
&=\mathbb{E}[Y\{a,Z(a),0\} \mid S = s]&&\text{(exclusion restriction)}\\
&=\mathbb{E}[Y^M(a,0) \mid S = s]&&\text{(composition)},
\end{align*}
Subtracting the two treatment-specific means gives
\begin{align*}
\mathrm{CDE}^Z_s=\mathrm{CDE}^M_s.
\end{align*}
\end{proof}

\appsubsection{Failure to identify CDE under elimination of unblinding}
\label{app:B-cde-nonidentification}

Assumptions \ref{assump:NDENIEidentification} and \ref{assump:NDENIEproxycompletenessbridge} identify $\mathrm{NDE}_s$ and $\mathrm{NIE}_s$, but do not generally identify $\mathrm{CDE}_s^M$, which we demonstrate with the following example.

Fix a trial and let $X$ be constant; we suppress both from the notation. Take four mutually independent uniform random variables on $(0,1)$, denoted $U_A,U_M,U_W,U_Z$, and set
\begin{align*}
A&=I(U_A<1/2),\qquad M= I(U_M<1/2),\\
W&=I(M = 0)I(U_W<b_0) + I(M = 1)I(U_W < b_1)\\
Z&=I(M = 0)I(U_Z<c_0) + I(M = 1)I(U_Z < c_1), \qquad Y=AW.
\end{align*}
Hence, $b_m = \mathrm{Pr}(W = 1 \mid M = m)$ and $c_m = \mathrm{Pr}(Z = 1 \mid M = m)$ for $m \in \{0,1\}$. Consider two models differing only in the proxy probabilities:
\[
\begin{array}{c|cc}
 & (b_0,b_1)&(c_0,c_1)\\\hline
\text{Model I}&(1/4,3/4)&(1/4,3/4)\\
\text{Model II}&(1/8,7/8)&(1/3,2/3).
\end{array}
\]
The structural equations imply $M(a)=M$ and $Y(a,m)=aI(U_W<b_m)$, which verifies the conditional exchangeability assumptions in Assumption \ref{assump:NDENIEidentification} in both models. Positivity also follows in both models as $\Pr(A = 1 \mid M) = \Pr(A = 1) = 1/2$. Moreover, one can verify the proxy independence conditions of Assumption \ref{assump:NDENIEproxycompletenessbridge} in both models. Since $c_1\ne c_0$ in both models, the matrix $[\Pr(M=m\mid Z=z,A=a)]_{z,m}$ can be shown by a simple application of Bayes' rule to be invertible, thereby establishing completeness. Lastly, the observed outcome bridge is simply $h_a(w)=aw$, because $Y=aW$ for individuals with $A = a$, and hence
\begin{align*}
\mathbb{E}[Y \mid Z, A = a] = \mathbb{E}[aW \mid Z, A = a]  = \mathbb{E}[h_a(W) \mid Z, A = a]
\end{align*}
Nevertheless, one can show that both models induce exactly the same observed distribution. Hence, while both models have
\begin{align*}
\mathrm{NDE} &= \mathbb{E}\{Y(1, M(0)) - Y(0, M(0))\}\\
&= \mathbb{E}\{ I(M = 0)I(U_W<b_0) + I(M = 1)I(U_W < b_1) \}\\
&= \frac{1}{2}b_0 + \frac{1}{2} b_1 = 1/2
\end{align*}
and
\begin{align*}
\mathrm{NIE} &= \mathbb{E}\{Y(1, M(1)) - Y(1, M(0))\}\\
&= 0,
\end{align*}
we have
\[
\mathrm{CDE}^M=\mathbb E\{Y(1,0)-Y(0,0)\}=b_0
=\begin{cases}1/4,&\text{Model I},\\1/8,&\text{Model II},\end{cases}
\]
which shows how Assumptions \ref{assump:NDENIEidentification} and \ref{assump:NDENIEproxycompletenessbridge} do not generally identify $\mathrm{CDE}_s^M$.

\appsubsection{Randomized interventional effects without intermediate confounding}
\label{app:B-interventional}

In this subsection and the next, all laws and assumptions concern a fixed trial, and we suppress conditioning on $S=s$.

Let $G_{a\mid X}$ be an externally randomized draw from the distribution of $M(a)$ conditional on $X$, independent of the recipient's counterfactual variables given $X$. Define
\begin{equation*}
\mathcal I(a',a)=\mathbb E\{Y(a',G_{a\mid X})\},\qquad
dF_{G_{a\mid X}\mid X}(m\mid x)=dF_{M(a)\mid X}(m\mid x).
\end{equation*}
From \citet{vanderweele2014effect}, the randomized interventional direct and indirect effects are
\begin{equation*}
\mathrm{RIDE}=\mathcal I(1,0)-\mathcal I(0,0),\qquad
\mathrm{RIIE}=\mathcal I(1,1)-\mathcal I(1,0).
\end{equation*}
\begin{assumption}
\label{assump:RIEnoL}
For each $a,a' \in \{0,1\}$,
\begin{enumerate}[(i)]
    \item Consistency: $M=M(a)$ if $A=a$ and 
$Y=Y(a,m)$ if $(A,M)=(a,m)$
    \item Conditional exchangeability. For $a \in \{0,1\}$,
    \begin{align*}
        &\{Y(a,m), M(a)\} \indep A\mid X,\\
        &Y(a,m) \indep M \mid A=a, X,
    \end{align*}
    \item Positivity: $0<\text{Pr}(A=1\mid M,X)<1$ almost surely
\end{enumerate}
\end{assumption}
\begin{assumption}
\label{assump:RIEnoLbridge}
Assume
\begin{enumerate}[(i)]
    \item $Y\indep Z\mid A,M,X, \qquad  W\indep (A,Z)\mid M,X$
    \item For each $a,x$ and every square-integrable $g$,
\[
\mathbb{E}\{g(M)\mid Z,A=a,X=x\}=0
\ \Longrightarrow\ g(M)=0\quad\text{almost surely},
\]
    \item There exists an integrable function $h_{a}(W,X)$ satisfying
\[
\mathbb{E}(Y\mid Z,A=a,X)
=\mathbb{E}\{h_a(W,X)\mid Z,A=a,X\}.
\]
\end{enumerate}
\end{assumption}

The above Assumptions \ref{assump:RIEnoL} and \ref{assump:RIEnoLbridge} are analogous to Assumptions \ref{assump:NDENIEidentification} and \ref{assump:NDENIEproxycompletenessbridge}, except: 1) they are no longer conditional on $S = s$, and 2) Assumption \ref{assump:RIEnoL}(ii) does not require the cross-world independence assumption $Y(a',m)\indep M(a)\mid X$.

\begin{theorem}[Outcome-bridge identification without treatment-induced confounding]
\label{thm:appB-ri}
Under Assumptions \ref{assump:RIEnoL} and \ref{assump:RIEnoLbridge},
\begin{align}
\mathcal I(a',a)
&=\iint h_{a'}(w,x)\,dF_{W\mid A,X}(w\mid a,x)\,dF_X(x)\notag
\end{align}
\end{theorem}

\begin{proof}
Let $\nu_{a'}(m,x)=\mathbb E(Y\mid A=a',M=m,X=x)$. By the randomization of $G_{a \mid X}$, followed by exchangeability and consistency:
\begin{align*}
\mathcal I(a',a)
&=\iint\mathbb E\{Y(a',m)\mid G_{a\mid X}=m,X=x\}
\,dF_{G_{a \mid X}\mid X}(m\mid x)\,dF_X(x)\\
&=\iint\mathbb E\{Y(a',m)\mid X=x\}
\,dF_{M(a)\mid X}(m\mid x)\,dF_X(x)\\
&=\iint\nu_{a'}(m,x)
\,dF_{M\mid A,X}(m\mid a,x)\,dF_X(x).
\end{align*}
Next define
\[
\kappa_{a'}(m,x)=\int h_{a'}(w,x)\,dF_{W\mid M,X}(w\mid m,x).
\]
The observed bridge equation and the two proxy restrictions imply
\begin{align*}
0&=\mathbb E\{Y-h_{a'}(W,x)\mid Z=z,A=a',X=x\}\\
&=\int\{\nu_{a'}(m,x)-\kappa_{a'}(m,x)\}
\,dF_{M\mid Z,A,X}(m\mid z,a',x).
\end{align*}
Completeness gives $\nu_{a'}(m,x)=\kappa_{a'}(m,x)$ for $F_{M\mid A,X}(\cdot\mid a',x)$-almost every $m$. In other words, for $g_{a',x}(m)=\nu_{a'}(m,x)-\kappa_{a'}(m,x)$,
\begin{align*}
    \text{Pr}\{ g_{a',x}(M) \neq 0 \mid A = a', X=x\} = 0
\end{align*}
for almost every $x$. By the positivity of the conditional mediator distribution, we have that
\begin{align*}
    &\text{Pr}\{ g_{a',x}(M) \neq 0 \mid A = a, X = x\}\\
    &= \mathbb{E}\left[ \frac{f_{M \mid A, X}(M \mid a,x)}{f_{M \mid A, X}(M \mid a',x)} I\{ g_{a',x}(M) \neq 0\} \mid A = a',X=x\right]=0
\end{align*}
and hence $\nu_{a'}(m,x)=\kappa_{a'}(m,x)$ for $F_{M\mid A,X}(\cdot\mid a,x)$-almost every $m$. Therefore
\begin{align*}
\mathcal I(a',a)
&=\iiint h_{a'}(w,x)\,dF_{W\mid M,X}(w\mid m,x)
\,dF_{M\mid A,X}(m\mid a,x)\,dF_X(x)\\
&=\iint h_{a'}(w,x)\,dF_{W\mid A,X}(w\mid a,x)\,dF_X(x).
\end{align*}
The last equality uses $W\indep A\mid M,X$.
\end{proof}

Under Assumptions \ref{assump:RIEnoL} and \ref{assump:RIEnoLbridge},
\[
\mathcal I(a,a)= \mathbb{E}\{Y(a)\},
\]
and the RIDE and RIIE sum to the ATE.

\appsubsection{Randomized interventional effects with an intermediate confounder}
\label{app:B-interventional-L}

Now let $L$ be a treatment-induced mediator-outcome confounder. In the following, note that the potential outcome $Y(a',m)$ leaves $L$ at its natural value under $a'$. As in the previous section, define
\[
\mathcal J(a',a)=\mathbb E\{Y(a',G_{a\mid X})\},
\]
where the external draw $G_{a\mid X}$ is still conditional only on $X$ and independent of the recipient's counterfactual variables given $X$, including $L(a')$. The confounder $L$ is observed and precedes $M$.

\begin{figure}[htbp]
\centering
\resizebox{\textwidth}{!}{%
\begin{tikzpicture}[
    >=Latex,
    nodebox/.style={draw, rounded corners=5pt, minimum width=1.45cm,
        minimum height=1.10cm, align=center, font=\large, inner sep=4pt},
    causal/.style={->, line width=1.1pt, draw=black},
    covariate/.style={->, line width=1.1pt, draw=black},
    direct/.style={->, line width=1.1pt, draw=black}
]
\node[nodebox] (A) at (0,0) {$A$};
\node[nodebox] (Z) at (3,0) {$Z$};
\node[nodebox] (M) at (6,0) {$M$};
\node[nodebox] (W) at (9,0) {$W$};
\node[nodebox] (Y) at (12,0) {$Y$};
\node[nodebox] (L) at (6,3) {$L$};
\node[nodebox] (X) at (6,-3) {$X$};

\draw[causal] (A) -- (Z);
\draw[causal] (Z) -- (M);
\draw[causal] (M) -- (W);
\draw[causal] (W) -- (Y);
\draw[causal] (A.south east) to[out=-25,in=205] (M.south west);
\draw[causal] (M.north east) to[out=25,in=155] (Y.north west);
\draw[causal] (A.north east) -- (L.south west);
\draw[causal] (L) -- (M);
\draw[causal] (L.south west) -- (Z.north);
\draw[causal] (L.south east) -- (Y.north west);
\draw[direct] (A.north) to[out=75,in=105,looseness=1.6] (Y.north);
\draw[covariate] (X.west) to[out=180,in=270] (A.south);
\draw[covariate] (X.north west) -- (Z.south);
\draw[covariate] (X) -- (M);
\draw[covariate] (X.north east) -- (W.south);
\draw[covariate] (X.east) to[out=0,in=270] (Y.south);
\end{tikzpicture}%
}
\caption{A DAG compatible with the conditional independence assumptions in
Assumptions \ref{assump:RIEwithL} and \ref{assump:RIEwithLbridge}.}
\label{fig:dag-intermediate-confounding}
\end{figure}
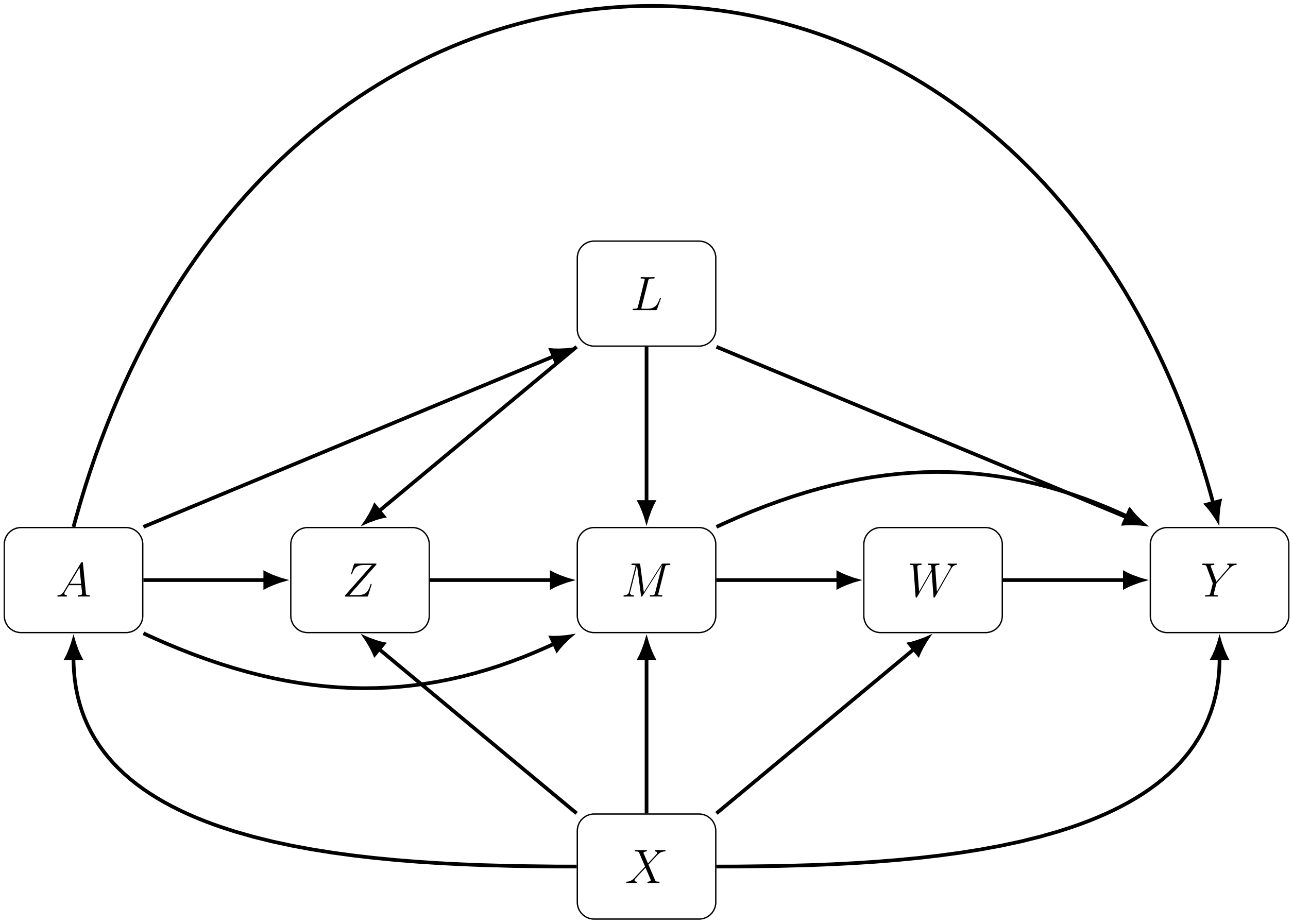

\begin{assumption}
\label{assump:RIEwithL}
For each $a,a'\in\{0,1\}$,
\begin{enumerate}[(i)]
    \item Consistency: $L=L(a)$ and $M=M(a)$ if $A=a$ and
$Y=Y(a,m)$ if $(A,M)=(a,m)$
    \item Conditional exchangeability. For $a \in \{0,1\}$,
    \begin{align*}
        &\{Y(a,m), M(a)\}\indep A\mid X,\notag\\
        &Y(a,m)\indep M\mid A=a,L,X
    \end{align*}
    \item Positivity: $0<\Pr(A=1\mid X)<1$ almost surely and for each $a \in \{0,1\}$ and $m\in\mathcal M$,
    \begin{equation*}
    f_{M\mid A,L,X}(m\mid A=a,L,X)>0 \quad \text{almost surely}.
    \end{equation*}
\end{enumerate}
\end{assumption}
\begin{assumption}
\label{assump:RIEwithLbridge}
Assume
\begin{enumerate}[(i)]
    \item $Y\indep Z\mid A,L,M,X,\qquad W\indep(A,L,Z)\mid M,X$,
    \item For each $a,l,x$ and every square-integrable $g$,
    \[
    \mathbb E\{g(M)\mid Z,A=a,L=l,X=x\}=0
    \ \Longrightarrow\ g(M)=0 \quad \text{almost surely},
    \]
    \item There exists an integrable function $h^L_a(W,L,X)$ satisfying
    \begin{equation*}
    \mathbb E(Y\mid Z,A=a,L,X)
    =\mathbb E\{h^L_a(W,L,X)\mid Z,A=a,L,X\}.
    \end{equation*}
\end{enumerate}
\end{assumption}
Assumptions \ref{assump:RIEwithL} and \ref{assump:RIEwithLbridge} parallel Assumptions \ref{assump:RIEnoL} and \ref{assump:RIEnoLbridge}, respectively, but allow for a treatment-induced mediator-outcome confounder $L$. Figure \ref{fig:dag-intermediate-confounding} gives one such DAG compatible with the conditional independence assumptions.

\begin{theorem}[Outcome-bridge identification with an observed intermediate confounder]
\label{thm:appB-riL}
Under Assumptions~\ref{assump:RIEwithL} and \ref{assump:RIEwithLbridge},
\begin{equation*}
\mathcal J(a',a)=\iiint h^L_{a'}(w,l,x)
\,dF_{W\mid A,X}(w\mid a,x)\,dF_{L\mid A,X}(l\mid a',x)\,dF_X(x).
\end{equation*}
\end{theorem}

\begin{proof}
Let $\nu^L_{a'}(l,m,x)=\mathbb E(Y\mid A=a',L=l,M=m,X=x)$. We have
\begin{align*}
	\mathcal J(a',a) &= \iint \mathbb{E} \{ Y(a',m) \mid G_{a \mid X} = m, X = x\} \, dF_{G_{a \mid X} \mid X}(m \mid x) \, dF_{X}(x)\\
	&= \iint \mathbb{E}\{ Y(a',m) \mid X = x \} \, dF_{M(a) \mid X} (m \mid x) \, dF_X (x)\\
	&= \iint \mathbb{E}\{ Y(a',m) \mid A = a', X = x \} \, dF_{M \mid A, X} (m \mid a, x) \, dF_X (x)\\
	&= \iiint \mathbb E\{Y(a',m)\mid A=a',L=l,X=x\}
\,dF_{L\mid A,X}(l\mid a',x)dF_{M \mid A, X} (m \mid a, x)\, dF_X (x)\\
&= \iiint \nu^L_{a'}(l,m,x)
\,dF_{L\mid A,X}(l\mid a',x)dF_{M \mid A, X} (m \mid a, x)\, dF_X (x)
\end{align*}
where the second equality follows from the randomization of $G_{a \mid X}$, the third equality follows from $Y(a',m) \indep A \mid X$, $M(a) \indep A \mid X$, and consistency, the fourth equality follows from iterated expectations, and the last equality follows from $Y(a',m) \indep M \mid A=a',L,X$ and consistency. Next, define
\[
\kappa^L_{a'}(l,m,x)=\int h^L_{a'}(w,l,x)\,dF_{W\mid M,X}(w\mid m,x).
\]
The observed bridge and proxy restrictions imply
\begin{align*}
0&=\mathbb E\{Y-h^L_{a'}(W,l,x)\mid Z=z,A=a',L=l,X=x\}\\
&=\int\{\nu^L_{a'}(l,m,x)-\kappa^L_{a'}(l,m,x)\}
\,dF_{M\mid Z,A,L,X}(m\mid z,a',l,x).
\end{align*}
Completeness implies $\nu^L_{a'}(l,m,x)=\kappa^L_{a'}(l,m,x)$ for $F_{M \mid L, A, X}(\cdot \mid l, a', x)$-almost every $m$. In other words, for $g_{a',l,x}^L(m) = \nu^L_{a'}(l,m,x) - \kappa^L_{a'}(l,m,x)$,
\begin{align*}
\mathrm{Pr}\{ g_{a',l,x}^L(M) \neq 0 \mid A = a', L = l, X = x\} = 0
\end{align*}
for almost every $(l,x)$. By positivity, we have that for almost every $(l,x)$,
\begin{align*}
&\Pr\{g^L_{a',l,x}(M)\ne0\mid A=a,X=x\}\\
&\quad=
\mathbb E\left[
\frac{f_{M\mid A,X}(M\mid a,x)}
     {f_{M\mid A,L,X}(M\mid a',l,x)}
I\{g^L_{a',l,x}(M)\ne0\}
\,\middle|\, A=a',L=l,X=x
\right]
=0
\end{align*}
and hence $\nu^L_{a'}(l,m,x)=\kappa^L_{a'}(l,m,x)$ for $F_{M \mid A, X}(\cdot \mid a, x)$-almost every $m$. Substituting $\kappa^L_{a'}(l,m,x)$ into the expression for $\mathcal J(a',a)$, invoking $W \indep A \mid M, X$, and applying iterated expectations yields the desired result.
\end{proof}

The corresponding effects are
\[
\mathrm{RIDE}_L=\mathcal J(1,0)-\mathcal J(0,0),\qquad
\mathrm{RIIE}_L=\mathcal J(1,1)-\mathcal J(1,0).
\]
In general $\mathcal J(a,a)\ne\mathbb E\{Y(a)\}$. The effects therefore sum to $\mathcal J(1,1)-\mathcal J(0,0)$, not necessarily to the ATE.

\appsubsection{Identification of natural effects standardized to a common covariate distribution}
\label{app:B-standardized}
In this section, we provide identification results for source-trial natural direct and indirect effects averaged over a target population's baseline covariate distribution when the mediator is unmeasured. 
Let $s,t \in \{0,1\}$ denote the trial index. Define
\begin{align}
\eta_{s,a',a}(x)
&=\mathbb E\{Y(a',M(a))\mid X=x,S=s\},\notag\\
\psi_{s\to t}(a',a)
&=\int\eta_{s,a',a}(x)\,dF_{X\mid S}(x\mid t).
\end{align}
Here, $\eta_{s,a',a}(x)$ is the conditional counterfactual mean in trial $s$, and $\psi_{s\to t}(a',a)$ denotes the result when marginalizing the conditional counterfactual mean in trial $s$ by the covariate distribution in trial $t$. Thus $\psi_{s\to s}(a',a)=\psi_s(a',a)$ in the main text.
\begin{assumption}
\label{assump:standardized-natural}
Trial positivity. Assume
\begin{align*}
0 < \Pr(S = 1 \mid X) < 1 \quad \text{almost surely}.
\end{align*}
\end{assumption}
\begin{theorem}
Under Assumptions \ref{assump:NDENIEidentification}, \ref{assump:NDENIEproxycompletenessbridge}, and \ref{assump:standardized-natural},
\begin{align*}
\psi_{s\to t}(a',a) = \iint h_{s,a'}(w,x)\,dF_{W\mid A,X,S}(w\mid a,x,s)\,dF_{X\mid S}(x\mid t).
\end{align*}
\end{theorem}
\begin{proof}
Closely following the proof of Theorem 1 from \citet{ghassami2025causal}, we have that
\[
\eta_{s,a',a}(x)
=\int h_{s,a'}(w,x)\,dF_{W\mid A,X,S}(w\mid a,x,s).
\]
Integrating with respect to $F_{X\mid S}(\cdot\mid t)$ gives the result.
\end{proof}

Define the conditional natural effects by
\[
\delta_s^{\mathrm{NDE}}(x)=\eta_{s,1,0}(x)-\eta_{s,0,0}(x),\qquad
\delta_s^{\mathrm{NIE}}(x)=\eta_{s,1,1}(x)-\eta_{s,1,0}(x).
\]
Their standardized versions are
\begin{align*}
\mathrm{NDE}_{s\to t}
&=\int\delta_s^{\mathrm{NDE}}(x)\,dF_{X\mid S}(x\mid t)
=\psi_{s\to t}(1,0)-\psi_{s\to t}(0,0),\\
\mathrm{NIE}_{s\to t}
&=\int\delta_s^{\mathrm{NIE}}(x)\,dF_{X\mid S}(x\mid t)
=\psi_{s\to t}(1,1)-\psi_{s\to t}(1,0).
\end{align*}
Hence, we also have the following decomposition:
\[
\mathrm{NDE}_{s\to t}+\mathrm{NIE}_{s\to t}
=\mathrm{ATE}_{s\to t}:=\psi_{s\to t}(1,1)-\psi_{s\to t}(0,0).
\]

\paragraph{Remarks.}
$\mathrm{NDE}_{s\to t}$ and $\mathrm{NIE}_{s\to t}$ may be interpreted as the target population's natural direct and indirect effects under additional transportability assumptions. For instance, if $\delta_s^{\mathrm{NDE}}(x)=\delta_t^{\mathrm{NDE}}(x)$, i.e., the conditional NDE is transportable across trials, then $\mathrm{NDE}_{s\to t}=\mathrm{NDE}_t$. An analogous statement would hold for the NIE.

Related work by \citet{rudolph2021transporting} and \citet{rudolph2022efficiently} identifies and estimates transported randomized interventional direct and indirect effects with measured mediators, allowing for observed treatment-induced mediator-outcome confounding. In particular, the target-mediator formulation in \citet{rudolph2022efficiently} combines the conditional outcome mean from the source population with the mediator, intermediate-confounder, and baseline-covariate distributions from the target population. Our formulation instead retains both the conditional outcome mean and the conditional mediator distribution from the source population, while averaging over the target population’s baseline covariate distribution. It also uses proxies to identify natural effects when the mediator is unmeasured.

These choices lead to different questions about reconciliation. To illustrate the distinction in our setting (without intermediate confounding), define
\[
D_{r,j;t}=\iint d_r(m,x)\,dF_{M \mid A, X, S}(m\mid 0, x, j)\,dF_{X\mid S}(x\mid t),
\]
where
\[
d_r(m,x) = \mathbb{E}[Y \mid A = 1, M = m, X = x, S = r] - \mathbb{E}[Y \mid A = 0, M = m, X = x, S = r].
\]
Here, $r$ indexes the population supplying the conditional outcome contrast, $j$ indexes the population supplying the placebo-arm mediator distribution, and $t$ indexes the population supplying the baseline covariate distribution. Our standardized effect is $\mathrm{NDE}_{s\to t}=D_{s,s;t}$, whereas the target population’s NDE is $\mathrm{NDE}_t=D_{t,t;t}$.

Under the identification assumptions, rejecting $D_{s,s;t}=D_{t,t;t}$ provides evidence that the conditional outcome contrast, the placebo-arm conditional mediator distribution, or both differ between populations. On the other hand, the direct effect estimand proposed in \citet{rudolph2021transporting} would parallel $D_{s,t;t}$. Rejecting $D_{s,t;t} = D_{t,t;t}$ would provide evidence specifically of a difference in the conditional outcome contrast across populations. One may also conceptualize the direct effect estimand $D_{t,s;t}$. Rejecting $D_{t,s;t}=D_{t,t;t}$ would provide evidence specifically of a difference in the placebo-arm conditional mediator distribution across populations. These comparisons clearly address distinct questions related to how components of the mediation pathway may differ across populations. We leave identification and estimation of $D_{s,t;t}$ and $D_{t,s;t}$ in our setting with an unmeasured mediator to future work.

\appsection{Appendix C}
\label{app:appendixC}
\appsubsection{Estimation and inference for the ATE, NDE, and NIE}
\label{app:C-effects}
We estimate the effects separately within each trial and suppress the trial index $s$. Let $O_i=(X_i,A_i,Z_i,W_i,Y_i)$, $i=1,\ldots,n$, denote the complete-case observations. We first discuss estimation of the following estimands presented in the main text:
\begin{align*}
	&\psi(1,1)=\mathbb E\{Y(1)\}\\
	&\psi(0,0)=\mathbb E\{Y(0)\}\\
	&\psi(1,0)=\mathbb E\{Y(1,M(0))\}
\end{align*}
Let
\begin{align*}
&p_a(x)=\Pr(A=a\mid X=x),\\
&\mu_a(x)=\mathbb E(Y\mid A=a,X=x),\\
&\eta_{1,0}(x)=\mathbb E\{h_1(W,X)\mid A=0,X=x\},
\end{align*}
where $a\in\{0,1\}$ and $h_1$ and $q_0$ are the outcome and treatment bridges defined in Appendix Section \ref{app:C-bridges}.

\paragraph{Cross-fitted estimators.}
Partition the observations into $K$ folds, and let $k_i$ denote observation $i$'s fold. A superscript $(-k_i)$ denotes a nuisance estimate trained using only observations outside that fold. The analyses in the main text use $K = 10$. Define
\begin{equation}
T_i(a)=\widehat\mu_a^{(-k_i)}(X_i)
+\frac{I(A_i=a)}{\widehat p_a^{(-k_i)}(X_i)}
\{Y_i-\widehat\mu_a^{(-k_i)}(X_i)\},\qquad a\in\{0,1\},
\end{equation}
and
\begin{align}
T_i(1,0)={}&\widehat\eta_{1,0}^{(-k_i)}(X_i)
+\frac{I(A_i=1)}{\widehat p_1^{(-k_i)}(X_i)}
\widehat q_0^{(-k_i)}(Z_i,X_i)
\{Y_i-\widehat h_1^{(-k_i)}(W_i,X_i)\}\notag\\
&+\frac{I(A_i=0)}{\widehat p_0^{(-k_i)}(X_i)}
\{\widehat h_1^{(-k_i)}(W_i,X_i)-\widehat\eta_{1,0}^{(-k_i)}(X_i)\},
\label{eq:appC-score}
\end{align}
where the expression in Equation \ref{eq:appC-score} follows from Corollary 1 of \citet{ghassami2025causal}. The resulting estimators are
\[
\widehat\psi(a,a)=\frac1n\sum_{i=1}^n T_i(a),\qquad
\widehat\psi(1,0)=\frac1n\sum_{i=1}^n T_i(1,0).
\]
The effect estimators are
\begin{align*}
\widehat{\mathrm{ATE}}&=\widehat\psi(1,1)-\widehat\psi(0,0),\\
\widehat{\mathrm{NDE}}&=\widehat\psi(1,0)-\widehat\psi(0,0),\\
\widehat{\mathrm{NIE}}&=\widehat\psi(1,1)-\widehat\psi(1,0).
\end{align*}
Swapping treatment labels in the expression for $T_i(1,0)$ yields $T_i(0,1)$ and the corresponding estimator of $\psi(0,1)$ for the alternative natural effect decomposition.

\paragraph{Nuisance estimation.}
We estimate $p_1$ by regressing $A$ on $X$, truncate predictions to $[10^{-6},1-10^{-6}]$, and set $\widehat p_0=1-\widehat p_1$. We estimate $\mu_a$ by regressing $Y$ on $X$ among training observations with $A=a$. For $\eta_{1,0}$, we first fit $h_1$ among observations with $A = 1$, predict $\widehat h_1(W_i,X_i)$ among observations with $A = 0$, and regress these predictions on $X$ among observations with $A = 0$. Estimation of the bridge functions $h_1$ and $q_0$ is discussed in more detail in Section \ref{app:C-bridges}.

While many different methods can be used to estimate $p_1$, $\mu_a$, and $\eta_{1,0}$, the analyses in the main text do so using SuperLearner with the library \texttt{SL.glm}, \texttt{SL.gam}, and \texttt{SL.mean} with 10-fold cross-validation. $\widehat p_1$ is estimated using the binomial family, and $\widehat \eta_{1,0}$ is estimated using the Gaussian family. $\widehat \mu_a$ is estimated using the binomial family when the outcome is binary and the Gaussian family when the outcome is gestational age in weeks.

\paragraph{Variance estimation and inference.}
Define
\[
\widehat\phi_{aa,i}=T_i(a)-\widehat\psi(a,a),\qquad
\widehat\phi_{10,i}=T_i(1,0)-\widehat\psi(1,0),
\]
and
\[
\widehat\phi_{\mathrm{ATE},i}=\widehat\phi_{11,i}-\widehat\phi_{00,i},\qquad
\widehat\phi_{\mathrm{NDE},i}=\widehat\phi_{10,i}-\widehat\phi_{00,i},\qquad
\widehat\phi_{\mathrm{NIE},i}=\widehat\phi_{11,i}-\widehat\phi_{10,i}.
\]
Let $\theta$ denote any of the parameters $\psi(0,0)$, $\psi(1,1)$, $\psi(1,0)$, $\mathrm{ATE}$, $\mathrm{NDE}$, or $\mathrm{NIE}$, and let $\widehat\theta$ be its estimator. Furthermore, let $\widehat\phi_{\theta,i}$ denote the corresponding estimated influence function value for observation $i$, as defined above; for example, when $\theta=\psi(1,0)$, $\widehat\phi_{\theta,i}=\widehat\phi_{10,i}$. Under regularity conditions,
\[
\sqrt n(\widehat\theta-\theta)\overset{d}{\longrightarrow}
N(0,\sigma_\theta^2),\qquad
\sigma_\theta^2=\mathbb E\{\phi_\theta(O)^2\}.
\]
We consider the following estimators
\[
\widehat\sigma_\theta^2=\frac1n\sum_{i=1}^n\widehat\phi_{\theta,i}^2,
\qquad
\widehat{\operatorname{Var}}(\widehat\theta)
=\frac{\widehat\sigma_\theta^2}{n}
=\frac1{n^2}\sum_{i=1}^n\widehat\phi_{\theta,i}^2.
\]
The estimated standard error is $\widehat{\operatorname{SE}}(\widehat\theta)=\sqrt{\widehat{\operatorname{Var}}(\widehat\theta)}$, giving the Wald 95\% confidence interval $\widehat\theta\pm1.96\,\widehat{\operatorname{SE}}(\widehat\theta)$. For contrasts between independent trials, i.e., for estimating $\mathrm{NDE}_1 - \mathrm{NDE}_0$, the two trial-specific variance estimates are added.

\appsubsection{Estimation of bridge functions}
\label{app:C-bridges}

In this section, we derive closed-form estimators for the bridge functions. We again work within one trial and suppress the trial index $s$. We focus on $\psi(1,0)=\mathbb E\{Y(1,M(0))\}$, as in the main text. The analogous procedure for $\psi(0,1)$ swaps treatment labels $0$ and $1$ everywhere.

The bridge equations are
\begin{align}
\mathbb{E}\{Y-h_1(W,X)\mid Z,A=1,X\}&=0,\\
\mathbb{E}\left\{\frac{I(A=1)}{p_1(X)}q_0(Z,X)
-\frac{I(A=0)}{p_0(X)}\,\middle|\,W,X\right\}&=0.
\end{align}
We use the adversarial objectives at the end of Section 4 of \citet{ghassami2025causal}.
We work in the finite-radius RKHS balls
\[
\mathcal H_{B_H}=\{h\in\mathcal H:\|h\|_{\mathcal H}\leq B_H\},
\qquad
\mathcal Q_{B_Q}=\{q\in\mathcal Q:\|q\|_{\mathcal Q}\leq B_Q\}.
\]
This formulation follows \citet[Appendix B.1]{olivas2025proximal}; see \citet{olivas2025source} for the associated source conditions and convergence analysis. Our main analysis uses $B_H=B_Q=50$.
Let $\mathbb E_n$ denote the empirical average over the training set and $\mathbb E_{n,1}$ the empirical average over the $A=1$ observations in the training set. Then
\begin{align}
\widehat h_1
&=\arg\min_{h\in\mathcal H_{B_H}}\sup_{f\in\mathcal F}
\Bigl[\mathbb E_{n,1}\bigl\{[Y-h(W,X)]f(Z,X)-f(Z,X)^2\bigr\}
\notag\\[-2pt]
&\hspace{4cm}-\lambda_F\|f\|_{\mathcal F}^2+\lambda_H\|h\|_{\mathcal H}^2\Bigr],
\label{eq:appC-h-adversarial}
\\
\widehat q_0
&=\arg\min_{q\in\mathcal Q_{B_Q}}\sup_{f\in\mathcal F}
\Biggl[\mathbb E_n\Biggl\{
\left[\frac{I(A=1)}{\widehat p_1(X)}q(Z,X)
-\frac{I(A=0)}{\widehat p_0(X)}\right]f(W,X)-f(W,X)^2\Biggr\}
\notag\\[-2pt]
&\hspace{4cm}-\lambda_F\|f\|_{\mathcal F}^2+\lambda_Q\|q\|_{\mathcal Q}^2\Biggr].
\label{eq:appC-g-adversarial}
\end{align}
Here $\mathcal H$ and $\mathcal Q$ are the bridge RKHSs; $\mathcal F$ denotes the respective test-function RKHS, on $(Z,X)$ for the outcome bridge and $(W,X)$ for the treatment bridge. The estimated propensity scores are held fixed in the treatment-bridge optimization. The RKHS derivation largely follows the approach in Appendix B.1 of \citet{olivas2025proximal}, adapted here for our specific bridge functions. All sample sizes below refer to the observations used for the particular fit, excluding any validation or outer evaluation fold.

\paragraph{Outcome bridge.}
We apply the closed-form outcome-bridge estimator in Appendix B.1 of \citet{olivas2025proximal} using the $m$ training observations with $A=1$. For $i = 1,\ldots,m$, let $U_i=(W_i,X_i)$, $V_i=(Z_i,X_i)$, and $y=(Y_1,\ldots,Y_m)^\top$. Let $k_H$ and $k_F$ be the kernels for the outcome bridge and its test function, respectively. Define the $m\times m$ Gram matrices
\[
K_{ij}=k_H(U_i,U_j),\qquad L_{ij}=k_F(V_i,V_j).
\]
Additionally, let
\[
\Gamma_m=\frac14\bigl(m^{-1}L+\lambda_F I_m\bigr)^{-1},
\]
where $I_m$ is the $m \times m$ identity matrix and $\lambda_H,\lambda_F>0$ are the bridge and test-function penalties. The estimator, before imposing the RKHS norm constraint discussed below, is
\begin{equation}
\widehat\alpha=(\Gamma_mLK+m^2\lambda_H I_m)^{-1}\Gamma_mLy,
\qquad
\widehat h_1(w,x)=\sum_{j=1}^m\widehat\alpha_j k_H\{(w,x),U_j\}.
\end{equation}

To enforce the prescribed norm bound, check whether $\widehat\alpha^\top K\widehat\alpha\leq B_H^2$. If so, the unconstrained solution above also solves the problem over $\mathcal H_{B_H}$. Otherwise, following \citet[Appendix B.1]{olivas2025proximal}, we replace $\widehat\alpha$ by
\[
\widehat\alpha(\xi_H)=\{\Gamma_mLK+m^2(\lambda_H+\xi_H)I_m\}^{-1}\Gamma_mLy,
\]
where the nonnegative Lagrange multiplier $\xi_H>0$ is chosen numerically to satisfy $\widehat\alpha(\xi_H)^\top K\widehat\alpha(\xi_H)=B_H^2$.

\paragraph{Treatment bridge.}
Let $n$ denote the number of observations in the training set. Define the $n\times n$ Gram matrices
\[
K_{ij}=k_Q(V_i,V_j),\qquad L_{ij}=k_F(U_i,U_j),
\]
where $k_Q$ and $k_F$ are the kernels for the treatment bridge and its test function, respectively. Let
\[
R=\operatorname{diag}\left\{\frac{I(A_i=1)}{\widehat p_1(X_i)}\right\}_{i=1}^n,
\qquad
u=\left(\frac{I(A_1=0)}{\widehat p_0(X_1)},\ldots,
\frac{I(A_n=0)}{\widehat p_0(X_n)}\right)^\top,
\]
and let $q_V=(q(V_1),\ldots,q(V_n))^\top$. The treatment-bridge objective in \eqref{eq:appC-g-adversarial} is
\begin{equation}
\sup_{f\in\mathcal F}\left[
\frac1n\sum_{i=1}^n\{(R_{ii}q(V_i)-u_i)f(U_i)-f(U_i)^2\}
-\lambda_F\|f\|_{\mathcal F}^2\right]+\lambda_Q\|q\|_{\mathcal Q}^2.
\end{equation}
Here $\lambda_F,\lambda_Q>0$, and the outer minimization is over $q\in\mathcal Q_{B_Q}$. We first derive the unconstrained coefficient solution and then enforce the norm bound.

By the generalized representer theorem, a maximizer of the inner problem has the form
\[
f(w,x)=\sum_{j=1}^n\beta_j k_F\{(w,x),U_j\},
\qquad \beta=(\beta_1,\ldots,\beta_n)^\top.
\]
Thus $f(U_i)=e_i^\top L\beta$, where $e_i$ is the $i$th standard basis vector in $\mathbb R^n$. The relevant terms are
\begin{align*}
\mathbb E_n\left[\frac{I(A=1)}{\widehat p_1(X)}q(Z,X)f(W,X)\right]
&=\frac1n\sum_{i=1}^n R_{ii}q(V_i)e_i^\top L\beta
=\frac1n q_V^\top RL\beta,\\
\mathbb E_n\left[\frac{I(A=0)}{\widehat p_0(X)}f(W,X)\right]
&=\frac1n\sum_{i=1}^n u_i e_i^\top L\beta
=\frac1n u^\top L\beta,\\
\mathbb E_n[f(W,X)^2]
&=\frac1n\sum_{i=1}^n\beta^\top Le_i e_i^\top L\beta
=\frac1n\beta^\top L^2\beta,\\
\|f\|_{\mathcal F}^2&=\beta^\top L\beta.
\end{align*}
Consequently, the inner objective is the quadratic function
\begin{align*}
&\frac1n q_V^\top RL\beta-\frac1n u^\top L\beta
-\frac1n\beta^\top L^2\beta-\lambda_F\beta^\top L\beta\\
&\qquad=\frac1n\beta^\top L(Rq_V-u)
-\beta^\top(n^{-1}L^2+\lambda_FL)\beta.
\end{align*}
with the solution
\[
\beta^*=\frac1{2n}(n^{-1}L+\lambda_F I_n)^{-1}(Rq_V-u).
\]
Define
\[
\zeta_n(q)=\frac1n(Rq_V-u),\qquad
\Gamma_n=\frac14(n^{-1}L+\lambda_F I_n)^{-1}.
\]
Substitution gives the maximized inner objective
\[
\zeta_n(q)^\top\Gamma_n L\zeta_n(q).
\]
The original optimization therefore reduces to
\[
\widehat q_0=\arg\min_{q\in\mathcal Q_{B_Q}}
\left\{\zeta_n(q)^\top\Gamma_n L\zeta_n(q)
+\lambda_Q\|q\|_{\mathcal Q}^2\right\}.
\]

Applying the representer theorem again, we obtain
\[
\widehat q_0(z,x)=\sum_{j=1}^n\gamma_j k_Q\{(z,x),V_j\},
\qquad \gamma=(\gamma_1,\ldots,\gamma_n)^\top.
\]
Then $q_V=K\gamma$. Substituting this representation into the objective function, we have
\begin{align*}
&n^{-2}(RK\gamma-u)^\top\Gamma_n L(RK\gamma-u)
+\lambda_Q\gamma^\top K\gamma\\
&\quad=n^{-2}\gamma^\top KR\Gamma_n LRK\gamma
-2n^{-2}\gamma^\top KR\Gamma_n Lu
+n^{-2}u^\top\Gamma_n Lu+\lambda_Q\gamma^\top K\gamma\\
&\quad=n^{-2}u^\top\Gamma_n Lu
-2n^{-2}\gamma^\top KR\Gamma_n Lu
+\gamma^\top\{n^{-2}KR\Gamma_n LRK+\lambda_QK\}\gamma.
\end{align*}
Here $\Gamma_n L$ is symmetric because $\Gamma_n$ is a function of $L$. Differentiating with respect to $\gamma$ gives
\[
-2n^{-2}KR\Gamma_n Lu
+2\{n^{-2}KR\Gamma_n LRK+\lambda_QK\}\gamma=0,
\]
or equivalently,
\[
K\{R\Gamma_n LRK\gamma+n^2\lambda_Q\gamma-R\Gamma_n Lu\}=0.
\]
A minimizer of the unconstrained convex quadratic objective is
\begin{equation}
\widehat\gamma=(R\Gamma_n LRK+n^2\lambda_Q I_n)^{-1}R\Gamma_n Lu,
\qquad
\widehat q_0(z,x)=\sum_{j=1}^n\widehat\gamma_j k_Q\{(z,x),V_j\}.
\end{equation}
For the treatment bridge, if $\widehat\gamma^\top K\widehat\gamma\leq B_Q^2$, the unconstrained solution also satisfies the prescribed RKHS norm bound. Otherwise, replace it by
\[
\widehat\gamma(\xi_Q)=\{R\Gamma_nLRK+n^2(\lambda_Q+\xi_Q)I_n\}^{-1}R\Gamma_nLu,
\]
where $\xi_Q>0$ is chosen numerically so that $\widehat\gamma(\xi_Q)^\top K\widehat\gamma(\xi_Q)=B_Q^2$.

\paragraph{Penalty scaling.}
The source-condition analysis of \citet[Theorems 4--5]{olivas2025source} relates bridge-estimation error to the regularization parameters, the critical radii of the relevant function classes, and the smoothness of the minimum RKHS-norm solutions. Under their source and regularity conditions, suitable penalty sequences yield $L_2$ consistency and bounds on the products of weak and strong bridge errors that enter asymptotic-normality arguments. Following \citet[Appendix B]{olivas2025proximal}, we scale the bridge penalty with the rate of the critical radius and the test-function penalty with order equal to the square of the critical radius. Specifically, letting $N$ denote the size of the training data ($m$ for the outcome bridge and $n$ for the treatment bridge) and for a Gaussian RKHS, we use
\begin{equation}
\lambda_F=\ell_{\mathrm{int}}\frac{\log N}{N},\qquad
\lambda_{\mathrm{bridge}}=\ell_{\mathrm{ext}}\sqrt{\frac{\log N}{N}},
\end{equation}
where $\lambda_{\mathrm{bridge}}$ denotes $\lambda_H$ or $\lambda_Q$.

\paragraph{Cross-validation.}
Estimation proceeds by 10-fold cross-validation. As the objective function for the treatment bridge requires $\widehat p_a$, for each split, we re-estimate $\widehat p_a$ using the split's training data and then obtain predictions for the split's training and validation data. In the main analyses, we consider the following grid for the various penalty parameters.
\[
\ell_{\mathrm{ext}}\in\{10^{-5},10^{-4},10^{-3},10^{-2},10^{-1}\},\qquad
\ell_{\mathrm{int}}\in\{10^{-1},1,10,100\}.
\]
We additionally tune the bandwidth $\sigma$ used for the Gaussian kernel $k(v,v')=\exp\{-\|v-v'\|^2/(2\sigma)\}$. Here $v$ and $v'$ denote kernel input vectors. For observation $i$, let $v_i^{\mathrm{ext}}$ and $v_i^{\mathrm{int}}$ denote the standardized inputs to the bridge and test-function kernels, respectively. For the outcome bridge, these are the standardized versions of $(W_i,X_i)$ and $(Z_i,X_i)$; for the treatment bridge, their roles are reversed. We use
\begin{align}
\label{eq:bandwidth}
\sigma_j
=s_j\{\operatorname{median}_{i<k}\|v_i^j-v_k^j\|\}^2/2,
\qquad j\in\{\mathrm{ext},\mathrm{int}\},
\end{align}
where the median is computed separately for each kernel using the training data of the current split. A zero median distance is replaced by the median of the positive distances; if no positive distances exist, the base bandwidth is set to one. We consider
\[
s_{\mathrm{ext}}\in\{1/4,1/2,1,2,4\},\qquad
s_{\mathrm{int}} = 1/4.
\]
In total, our main analyses considered 100 candidate combinations per bridge.

Let $n_{\mathrm{val}}$ denote the size of the validation subset in each split, and let $r_h=y-\widehat h$ or $r_q=R\widehat q-u$. Here, we again emphasize that $\widehat p_a$ in $R,u$ is estimated using a given split's training data. Define
\[
\Gamma_{\mathrm{val}}=\frac14\{n_{\mathrm{val}}^{-1}L_{\mathrm{val}}+\log(n_{\mathrm{val}})/n_{\mathrm{val}}\,I_{n_{\mathrm{val}}}\}^{-1},
\]
where the validation test-function Gram matrix $L_{\mathrm{val}}$ uses the median distance among validation observations, scaled in the main analyses by $1/4$ (i.e., parallel to choosing $s_j = 1/4$ in Equation (\ref{eq:bandwidth})). The validation risks are
\begin{equation}
\widehat{\mathcal R}_h=\frac1{n_{\mathrm{val}}^2} r_h^\top\Gamma_{\mathrm{val}}L_{\mathrm{val}}r_h,
\qquad
\widehat{\mathcal R}_q=\frac1{n_{\mathrm{val}}^2} r_q^\top\Gamma_{\mathrm{val}}L_{\mathrm{val}}r_q.
\end{equation}
Final average risks for a candidate tuning parameter are computed as the unweighted mean of the finite validation risks across splits. After selecting the candidate tuning parameters with the minimum average risk, the bridge function is refitted on both the training and validation data, recomputing base bandwidths on the new sample, and rescaling penalties using the new sample size.

\appsubsection{Estimation and inference for the CDE}
\label{app:C-cde-estimation}

We estimate the CDE separately within each trial and suppress the trial index $s$. For this section, let $O=(X,A,Z,Y)$ and define
\[
p_a(x)=\Pr(A=a\mid X=x),\qquad
r_a(x)=\Pr(Z=0\mid A=a,X=x),\qquad
e_a(x)=p_a(x)r_a(x),
\]
\[
m_a(x)=\mathbb E(Y\mid A=a,Z=0,X=x),\qquad
\theta_a^Z=\mathbb E\{Y^Z(a,0)\},\qquad
\mathrm{CDE}^Z=\theta_1^Z-\theta_0^Z.
\]
Under the identification assumptions in Appendix Section \ref{app:B-cde-identification}, within each trial, $\theta_a^Z:=\theta_s^{Z}(a,0)=\mathbb E\{m_a(X)\}$. In the nonparametric observed-data model, its efficient influence function \citep{van2003unified} is
\begin{equation}
\phi_{a}(O)
=\frac{I(A=a,Z=0)}{e_{a}(X)}\{Y-m_{a}(X)\}
+m_{a}(X)-\theta^Z_{a}.
\label{eq:appB-cde-eif-mean}
\end{equation}
The efficient influence function for $\mathrm{CDE}^Z$ is $\phi_{1}(O)-\phi_{0}(O)$. Estimation proceeds using standard 10-fold cross-fitting. Use superscript $(-k_i)$ for nuisance estimates fitted without the fold containing observation $i$. The cross-fit estimator is:
\begin{align}
\widehat\theta^Z_{a}
&=\frac1{n}\sum_{i=1}^n\left[
\widehat m_{a}^{(-k_i)}(X_i)
+\frac{I(A_i=a,Z_i=0)}{\widehat p_{a}^{(-k_i)}(X_i)
\widehat r_{a}^{(-k_i)}(X_i)}
\{Y_i-\widehat m_{a}^{(-k_i)}(X_i)\}\right],\\
\widehat{\mathrm{CDE}}^Z
&=\widehat\theta^Z_{1}-\widehat\theta^Z_{0}.\notag
\end{align}
Nuisance functions are estimated using SuperLearner, though other methods may also be considered. We consider a library of \texttt{SL.glm}, \texttt{SL.gam}, and \texttt{SL.mean} with 10-fold cross-validation. In particular, we estimate $p_{a}$ by regressing treatment assignment on baseline covariates, $r_{a}$ by regressing $I(Z=0)$ on baseline covariates within treatment arm $a$, and $m_{a}$ by regressing the outcome on baseline covariates among participants with $A=a,Z=0$. Predictions for $\widehat p_{a}$ and $\widehat r_{a}$ are truncated to $[10^{-6},1-10^{-6}]$.

Let $\widehat\phi_{a,i}$ denote Equation \eqref{eq:appB-cde-eif-mean} with cross-fitted nuisance estimates and $\widehat\theta^Z_{a}$ substituted, and let $\widehat\phi_{\mathrm{CDE},i}=\widehat\phi_{1,i}-\widehat\phi_{0,i}$. Under regularity conditions, the variance and a 95\% confidence interval can be estimated by
\[
\widehat V=\frac1{n^2}\sum_{i=1}^n\widehat\phi_{\mathrm{CDE},i}^{\,2},
\qquad
\widehat{\mathrm{CDE}}^Z\ \pm\ 1.96\sqrt{\widehat V}.
\]
For contrasts involving the estimated CDE across different trials, the estimated variances of the individual estimators may be added.

\appsubsection{Estimation and inference for standardized natural effects}
\label{app:C-standardized}

In this section, we discuss estimation and inference for the standardized natural effect estimands proposed in Section \ref{app:B-standardized}. Let $n$ denote the total sample size, and
\[
n_r=\sum_{i=1}^n I(S_i=r),\qquad r\in\{0,1\},
\]
with $n_0+n_1=n$. Define
\begin{align*}
\rho_s(x)=\Pr(S=s\mid X=x),
\end{align*}
We propose the following cross-fitted estimator for $\psi_{s\to t}(a',a)$:
\begin{align*}
\widehat\psi_{s\to t}(a',a)
=\frac1{n_t}\sum_{i=1}^n\Bigg[&
I(S_i=t)\widehat\eta_{s,a',a}^{(-k_i)}(X_i)\notag\\
&+I(S_i=s)\frac{\widehat\rho_t^{(-k_i)}(X_i)}{\widehat\rho_s^{(-k_i)}(X_i)}
\Bigg\{\frac{I(A_i=a')}{\widehat p_{s,a'}^{(-k_i)}(X_i)}
\widehat q_{s,a}^{(-k_i)}(Z_i,X_i)\notag\\
&\hspace{25mm}\times\{Y_i-\widehat h_{s,a'}^{(-k_i)}(W_i,X_i)\}\notag\\
&\hspace{13mm}+\frac{I(A_i=a)}{\widehat p_{s,a}^{(-k_i)}(X_i)}
\{\widehat h_{s,a'}^{(-k_i)}(W_i,X_i)-\widehat\eta_{s,a',a}^{(-k_i)}(X_i)\}
\Bigg\}\Bigg].
\end{align*}
When $a'=a$, the estimator simplifies to the standard transportability estimator proposed in \citet{dahabreh2020extending} that does not require estimation of bridge functions:
\begin{align*}
\widehat\psi_{s\to t}(a,a)
=\frac1{n_t}\sum_{i=1}^n\Bigg[&
I(S_i=t)\widehat\mu_{s,a}^{(-k_i)}(X_i)\notag\\
&+I(S_i=s)\frac{\widehat\rho_t^{(-k_i)}(X_i)}{\widehat\rho_s^{(-k_i)}(X_i)}
\frac{I(A_i=a)}{\widehat p_{s,a}^{(-k_i)}(X_i)}
\{Y_i-\widehat\mu_{s,a}^{(-k_i)}(X_i)\}\Bigg].
\end{align*}
In the above expressions, we denote estimates trained outside fold $k_i$ by superscript $(-k_i)$. Our analyses use 10-fold cross-fitting. Nuisance functions are fit according to Appendix Sections \ref{app:C-effects} and \ref{app:C-bridges}, with $\widehat \rho_1$ being estimated using SuperLearner (using library \texttt{SL.glm}, \texttt{SL.gam}, and \texttt{SL.mean} with 10-fold cross-validation and the binomial family) by regressing $S$ on $X$ in each split, truncating predictions to $[10^{-6},1-10^{-6}]$, and letting $\widehat \rho_0 = 1 - \widehat \rho_1$.

We propose the following estimators of the standardized natural effects:
\begin{align*}
\widehat{\mathrm{NDE}}_{s\to t}
&=\widehat\psi_{s\to t}(1,0)-\widehat\psi_{s\to t}(0,0),\\
\widehat{\mathrm{NIE}}_{s\to t}
&=\widehat\psi_{s\to t}(1,1)-\widehat\psi_{s\to t}(1,0).
\end{align*}
For $a'\ne a$, we use the observation-level estimated influence function values
\begin{align*}
\widehat\phi_{s\to t,i}^{a',a}
=\frac{n}{n_t}\Bigg[&I(S_i=t)
\{\widehat\eta_{s,a',a}^{(-k_i)}(X_i)-\widehat\psi_{s\to t}(a',a)\}\\
&+I(S_i=s)\frac{\widehat\rho_t^{(-k_i)}(X_i)}{\widehat\rho_s^{(-k_i)}(X_i)}
\Bigg\{\frac{I(A_i=a')}{\widehat p_{s,a'}^{(-k_i)}(X_i)}
\widehat q_{s,a}^{(-k_i)}(Z_i,X_i)\\
&\hspace{25mm}\times\{Y_i-\widehat h_{s,a'}^{(-k_i)}(W_i,X_i)\}\\
&\hspace{13mm}+\frac{I(A_i=a)}{\widehat p_{s,a}^{(-k_i)}(X_i)}
\{\widehat h_{s,a'}^{(-k_i)}(W_i,X_i)-\widehat\eta_{s,a',a}^{(-k_i)}(X_i)\}
\Bigg\}\Bigg].
\end{align*}
For $a'=a$, the corresponding values are
\begin{align*}
\widehat\phi_{s\to t,i}^{a,a}
=\frac{n}{n_t}\Bigg[&I(S_i=t)
\{\widehat\mu_{s,a}^{(-k_i)}(X_i)-\widehat\psi_{s\to t}(a,a)\}\\
&+I(S_i=s)\frac{\widehat\rho_t^{(-k_i)}(X_i)}{\widehat\rho_s^{(-k_i)}(X_i)}
\frac{I(A_i=a)}{\widehat p_{s,a}^{(-k_i)}(X_i)}
\{Y_i-\widehat\mu_{s,a}^{(-k_i)}(X_i)\}\Bigg].
\end{align*}
For the standardized natural effects, use
\begin{align*}
\widehat\phi_{s\to t,i}^{\mathrm{NDE}}
&=\widehat\phi_{s\to t,i}^{1,0}-\widehat\phi_{s\to t,i}^{0,0},\\
\widehat\phi_{s\to t,i}^{\mathrm{NIE}}
&=\widehat\phi_{s\to t,i}^{1,1}-\widehat\phi_{s\to t,i}^{1,0}.
\end{align*}
For any of these estimands, denoted generically by $\vartheta$, we propose the following variance estimator:
\[
\widehat{\operatorname{Var}}(\widehat\vartheta)
=\frac1{n^2}\sum_{i=1}^n
\widehat\phi_{\vartheta,i}^{\,2}.
\]
95\% Wald confidence intervals are given by $\widehat\vartheta\pm1.96\sqrt{\widehat{\operatorname{Var}}(\widehat\vartheta)}$. We leave a more precise asymptotic characterization of the proposed estimators in this section for future work.

\appsubsection{Estimation of randomized interventional effects with an intermediate confounder}
\label{app:C-interventional-L}

In this section, we propose estimators for the randomized interventional effects identified in Appendix \ref{app:B-interventional-L}. We work within one trial and suppress the trial index $s$. Let $O_i=(X_i,A_i,L_i,Z_i,W_i,Y_i)$, $i=1,\ldots,n$, denote the observed data, where $L$ is a measured treatment-induced mediator-outcome confounder. We adapt the multiply robust one-step estimator of \citet{diaz2021nonparametric}, which is based on the efficient influence function when the mediator is measured, by replacing the outcome regression and mediator density ratio with their respective bridge functions.

Let $p_a(x)=\Pr(A=a\mid X=x)$. For each $a',a\in\{0,1\}$, consider square-integrable bridge functions satisfying the two integral equations
\begin{align}
\mathbb E\{Y-h^L_{a'}(W,L,X)\mid Z,A=a',L,X\}&=0,
\label{eq:appC-riL-h-bridge}\\
\mathbb E\{q^L_{a',a}(Z,L,X)\mid W,A=a',L,X\}
&=\frac{f_{W\mid A,X}(W\mid a,X)}
{f_{W\mid A,L,X}(W\mid a',L,X)}.
\label{eq:appC-riL-q-bridge}
\end{align}
Define
\begin{align*}
u^L_{a',a}(l,x)&=\int h^L_{a'}(w,l,x)\,dF_{W\mid A,X}(w\mid a,x),\\
v^L_{a'}(w,x)&=\int h^L_{a'}(w,l,x)\,dF_{L\mid A,X}(l\mid a',x),\\
\eta^L_{a',a}(x)&=\int u^L_{a',a}(l,x)\,dF_{L\mid A,X}(l\mid a',x)\\
&=\int v^L_{a'}(w,x)\,dF_{W\mid A,X}(w\mid a,x).
\end{align*}
By Theorem \ref{thm:appB-riL}, $\mathcal J(a',a)=\mathbb E\{\eta^L_{a',a}(X)\}$. Following Appendix \ref{app:C-effects}, use $K$-fold cross-fitting and denote nuisance estimates fitted without observation $i$'s fold by superscript $(-k_i)$. Let
\begin{align*}
T_i^L(a',a)={}&\widehat\eta^{L,(-k_i)}_{a',a}(X_i)\\
&+\frac{I(A_i=a')}{\widehat p_{a'}^{(-k_i)}(X_i)}
\widehat q^{L,(-k_i)}_{a',a}(Z_i,L_i,X_i)
\{Y_i-\widehat h^{L,(-k_i)}_{a'}(W_i,L_i,X_i)\}\\
&+\frac{I(A_i=a')}{\widehat p_{a'}^{(-k_i)}(X_i)}
\{\widehat u^{L,(-k_i)}_{a',a}(L_i,X_i)-\widehat\eta^{L,(-k_i)}_{a',a}(X_i)\}\\
&+\frac{I(A_i=a)}{\widehat p_a^{(-k_i)}(X_i)}
\{\widehat v^{L,(-k_i)}_{a'}(W_i,X_i)-\widehat\eta^{L,(-k_i)}_{a',a}(X_i)\},\\
\widehat{\mathcal J}(a',a)={}&\frac1n\sum_{i=1}^n T_i^L(a',a).
\end{align*}
The corresponding effect estimators are
\begin{align*}
\widehat{\mathrm{RIDE}}_L&=\widehat{\mathcal J}(1,0)-\widehat{\mathcal J}(0,0),\\
\widehat{\mathrm{RIIE}}_L&=\widehat{\mathcal J}(1,1)-\widehat{\mathcal J}(1,0).
\end{align*}
Direct estimation based on the preceding expressions generally requires estimating conditional densities involving $W$ and $L$ in order to estimate $u_{a',a}^L$, $v_{a',a}^L$, $\eta_{a',a}^L$, and $q_{a',a}^L$. This can be challenging when $W$ and $L$ are continuous or high-dimensional. Following \citet{diaz2021nonparametric}, we consider an alternative formulation better suited to continuous or high-dimensional $W$ and low-dimensional $L$. As also discussed in \citet{diaz2021nonparametric}, a complementary formulation for low-dimensional $W$ and high-dimensional $L$ may also be possible.

Note that $u_{a',a}^L$ and $\eta_{a',a}^L$ can equivalently be expressed as:
\begin{align}
u^L_{a',a}(l,x)
&=\mathbb E\{h^L_{a'}(W,L,X)q^L_{a',a}(Z,L,X)\mid A=a',L=l,X=x\},
\label{eq:appC-riL-u-regression}\\
\eta^L_{a',a}(x)
&=\mathbb E\{v^L_{a'}(W,X)\mid A=a,X=x\}.
\label{eq:appC-riL-eta-regression}
\end{align}
This motivates the following estimation strategy within each training fold. Given estimates of both bridge functions, regress the pseudo-outcome $\widehat h^L_{a'}(W,L,X)\widehat q^L_{a',a}(Z,L,X)$ on $(L,X)$ among participants with $A=a'$ to estimate $u^L_{a',a}$. Estimate $v^L_{a'}$ by
\[
\widehat v^L_{a'}(w,x)
=\int\widehat h^L_{a'}(w,l,x)\,d\widehat F_{L\mid A,X}(l\mid a',x),
\]
which can directly be evaluated by a summation when $L$ is low-dimensional. Regress $\widehat v^L_{a'}(W,X)$ on $X$ among participants with $A=a$ to estimate $\eta^L_{a',a}$. The propensity score $p_a$ can likewise be estimated or replaced by known randomization probabilities.

Estimation of $q^L_{a',a}$ using the current moment equation in Equation \eqref{eq:appC-riL-q-bridge} as presented would also require estimation of conditional densities involving $W$. Following Proposition 1 of \citet{ghassami2025causal}, we have the following equivalent conditional moment representation instead in terms of conditional densities involving $L$.

\begin{proposition}
\label{prop:appC-riL-q-moment}
The function $q^L_{a',a}$ solves Equation \eqref{eq:appC-riL-q-bridge} if and only if it solves
\begin{equation}
\begin{aligned}
\mathbb E\Biggl\{&
\frac{I(A=a')}{p_{a'}(X)}q^L_{a',a}(Z,L,X)\\
&-\frac{I(A=a)}{p_a(X)}\frac{f_{L\mid A,X}(L\mid a',X)}{f_{L\mid W,A,X}(L\mid W,a,X)}
\,\Biggm|\,W,L,X\Biggr\}=0.
\end{aligned}
\label{eq:appC-riL-q-moment}
\end{equation}
\end{proposition}
\begin{proof}
By Bayes' rule,
\begin{align*}
&\frac{f_{W\mid A,X}(W\mid a,X)}{f_{W\mid A,L,X}(W\mid a',L,X)}\\
&\qquad=\frac{\Pr(A=a\mid W,L,X)}{\Pr(A=a'\mid W,L,X)}
\frac{p_{a'}(X)}{p_a(X)}\frac{f_{L\mid A,X}(L\mid a',X)}{f_{L\mid W,A,X}(L\mid W,a,X)}.
\end{align*}
Consequently, Equation \eqref{eq:appC-riL-q-bridge} is equivalent to
\begin{align*}
&\frac{\Pr(A=a'\mid W,L,X)}{p_{a'}(X)}
\mathbb E\{q^L_{a',a}(Z,L,X)\mid W,A=a',L,X\}\\
&\qquad=\frac{\Pr(A=a\mid W,L,X)}{p_a(X)}\frac{f_{L\mid A,X}(L\mid a',X)}{f_{L\mid W,A,X}(L\mid W,a,X)},
\end{align*}
which, by iterated expectations, is equivalent to Equation \eqref{eq:appC-riL-q-moment}.
\end{proof}

Proposition \ref{prop:appC-riL-q-moment} motivates an adversarial estimation approach analogous to Appendix \ref{app:C-bridges}. Let $\mathcal H_{B_H}$ and $\mathcal Q_{B_Q}$ be finite-radius RKHS balls on $(W,L,X)$ and $(Z,L,X)$, respectively, and let $\mathcal F_H$ and $\mathcal F_Q$ be test-function RKHSs on $(Z,L,X)$ and $(W,L,X)$. Let $\mathbb E_n$ denote the empirical average over the training set and $\mathbb E_{n,a'}$ denote the empirical average over the training set among observations with $A=a'$. For the outcome bridge, consider
\begin{align}
\widehat h^L_{a'}
&=\arg\min_{h\in\mathcal H_{B_H}}\sup_{t\in\mathcal F_H}
\Bigl[\mathbb E_{n,a'}\bigl\{[Y-h(W,L,X)]t(Z,L,X)-t(Z,L,X)^2\bigr\}
\notag\\[-2pt]
&\hspace{3cm}-\lambda_{F_H}\|t\|_{\mathcal F_H}^2+\lambda_H\|h\|_{\mathcal H}^2\Bigr].
\label{eq:appC-riL-h-adversarial}
\end{align}
For the treatment bridge, consider
\begin{align}
\widehat q^L_{a',a}
&=\arg\min_{q\in\mathcal Q_{B_Q}}\sup_{t\in\mathcal F_Q}
\Biggl[\mathbb E_n\Biggl\{
\left[\frac{I(A=a')}{\widehat p_{a'}(X)}q(Z,L,X)\right.
\notag\\[-2pt]
&\hspace{2cm}\left.-\frac{I(A=a)}{\widehat p_a(X)}\frac{\widehat f_{L\mid A,X}(L\mid a',X)}{\widehat f_{L\mid W,A,X}(L\mid W,a,X)}\right]t(W,L,X)
-t(W,L,X)^2\Biggr\}
\notag\\[-2pt]
&\hspace{2cm}-\lambda_{F_Q}\|t\|_{\mathcal F_Q}^2+\lambda_Q\|q\|_{\mathcal Q}^2\Biggr].
\label{eq:appC-riL-q-adversarial}
\end{align}
These objectives are similar to those in Equations \eqref{eq:appC-h-adversarial} and \eqref{eq:appC-g-adversarial} in Appendix \ref{app:C-bridges}, except with the conditional density ratio of $L$ entering into the objective function for the treatment bridge function. This suggests that a similar estimation strategy for the bridge functions may be used.

We leave a more precise asymptotic characterization of the proposed estimators in this section for future work.


\end{document}